\documentclass[conference]{IEEEtran}
\pdfoutput=1
\IEEEoverridecommandlockouts
\usepackage{cite}
\usepackage{amsmath,amssymb,amsfonts}
\usepackage{algorithmic}
\usepackage{graphicx}
\usepackage{textcomp}
\usepackage{xcolor}
\usepackage{tcolorbox}
\usepackage{array}
\usepackage{caption}
\usepackage{subcaption}
\usepackage{multirow}
\usepackage{enumitem}
\usepackage{xurl}
\newcolumntype{L}[1]{>{\raggedright\let\newline\\\arraybackslash\hspace{0pt}}m{#1}}
\newcolumntype{C}[1]{>{\centering\let\newline\\\arraybackslash\hspace{0pt}}m{#1}}
\newcolumntype{R}[1]{>{\raggedleft\let\newline\\\arraybackslash\hspace{0pt}}m{#1}}
\usepackage[font=small,skip=1pt]{caption}
\usepackage{verbatim}
\def\BibTeX{{\rm B\kern-.05em{\sc i\kern-.025em b}\kern-.08em
    T\kern-.1667em\lower.7ex\hbox{E}\kern-.125emX}}

\usepackage[resetlabels,labeled]{multibib}
\newcites{B}{References of blog articles}

\begin{document}
\title{Where and What do Software Architects blog? \\ {\large An Exploratory Study on Architectural Knowledge in Blogs, and their Relevance to Design Steps}}

\author{\IEEEauthorblockN{
Mohamed Soliman\IEEEauthorrefmark{1},
Kirsten Gericke\IEEEauthorrefmark{1}
and
Paris Avgeriou\IEEEauthorrefmark{1}}
\IEEEauthorblockA{\IEEEauthorrefmark{1}
Bernoulli Institute for Mathematics, Computer Science and Artificial Intelligence \\
University of Groningen, Groningen, The Netherlands \\
\{m.a.m.soliman, p.avgeriou\}@rug.nl, kirstykromhout7@gmail.com}
}

\maketitle
\begin{abstract}
Software engineers share their architectural knowledge (AK) in different places on the Web. Recent studies show that architectural blogs contain the most relevant AK, which can help software engineers to make design steps. Nevertheless, we know little about blogs, and specifically architectural blogs, where software engineers share their AK. In this paper, we conduct an exploratory study on architectural blogs to explore their types, topics, and their AK. Moreover, we determine the relevance of architectural blogs to make design steps. Our results support researchers and practitioners to find and re-use AK from blogs.
\end{abstract}


\section{Introduction}
\label{sec:introduction}

Software engineers need \textit{architectural knowledge} (AK) \cite{Kruchten2006} %(e.g. knowledge on technologies and their risks \cite{SolimanWicsa2015}) 
to make reasonable \textit{architectural design decisions} \cite{Jansen2005}. For example, when software engineers work on component design (i.e. the structure and behavior of the system components), they often rely on their AK from previous projects, or from discussions with experienced architects. To understand the nature of this AK, researchers have explored various \textit{AK concepts} \cite{Soliman2021ExploringKnowledge}, such as architectural solutions (e.g. patterns \cite{BuschmannHenneySchmidt07a} and technologies \cite{SolimanWicsa2015}), as well as constraints, benefits and drawbacks of solutions \cite{Tang2007a}. %These AK concepts constitute the building blocks of AK, which software engineers need to make design decisions.

Software engineers make design decisions in consecutive steps, and for each step, they need different AK concepts \cite{Soliman2018ImprovingCommunities}. To exemplify this, let us consider an example of such a step-wise process, the Attribute Driven Design (ADD) \cite{KazmanDesigningSoftware2016}, the most well-established architecture design process (see Section \ref{sec:background}). In the \textit{identify design concept} step of ADD, AK concepts such as architectural solutions and quality attributes (e.g. performance and security) are considered. In the \textit{select design concept} step, AK concepts such as benefits and drawbacks between alternative solutions are considered. In the \textit{instantiate architecture elements} step, AK concepts such as components and connectors are considered.
%For instance, to decide on a Big Data technology, software engineers might consider three AK concepts: \textit{alternative technologies}, the \textit{benefits and drawbacks} of technologies (e.g. regarding performance and security), and \textit{constraints} such as team skills and organizational policies.  For instance, 
While AK is critical for those design steps, it is rather challenging to find AK. For instance, software engineers struggle to find design decisions from architectural documents, because design decisions are not systematically documented and formalized \cite{Manteuffel2018AnProjects}. 

To address this problem, researchers have recently explored AK in software repositories \cite{Hassan2008TheRepositories} such as issue tracking systems \cite{Bhat2017AutomaticApproach}, and mailing lists \cite{Li2020AutomaticList}, as well as in certain web resources such as Stack Overflow \cite{Bi2021MiningOverflow} and technology documentations \cite{GortonICSA2017}. These contributions provide concrete datasets, AK ontologies \cite{Soliman2017DevelopingCommunities}, and search approaches \cite{Soliman2018ImprovingCommunities} that can help researchers and practitioners to find AK.
Moreover, there is evidence that software engineers share their AK in technical articles within blogs (i.e. \textit{architectural blogs}) \cite{Soliman2021ExploringKnowledge}. Such blogs showed to be the most relevant source on the Web to perform architectural tasks, and the richest source of AK concepts, compared to forums (e.g. Stack Overflow), source code repositories and technology documentations \cite{Soliman2021ExploringKnowledge}.

Despite the value of architectural blogs in finding AK, software engineering researchers rarely explored blogs (e.g. \cite{WalidBlogs}), and did not specifically explore architectural blogs and their contained AK. Moreover, architectural blogs are quite unstructured and disorganized \cite{Williams2017TowardResearch}: they are scattered on the Web among thousands of websites and their \textit{type} varies  from personal blogs to blogs hosted by software companies and communities. This makes it hard to understand which blogs one should look for AK in. %traditional web search engines (e.g. Google) do not classify blogs and their types.
Furthermore, architectural blogs discuss multiple \textit{topics} (e.g. comparing solutions), each involving different AK concepts. However, blog articles are not classified (e.g. using tags) \cite{WalidBlogs}, which makes it challenging to search for and subsequently reuse their contained AK concepts. Therefore, in this paper, we aim to \textit{explore the types and topics of architectural blogs, as retrieved by Web search engines, and their relevance for software engineers during certain design steps}. This allows us to provide an overview of architectural blogs, which can be used by researchers and practitioners to find and re-use AK from architectural blogs.

To this end, we empirically analyzed the dataset of architectural blogs from Soliman et al. \cite{Soliman2021ExploringKnowledge}, and applied grounded theory to explore the types and topics of architectural blogs. Moreover, we applied the most popular topic modeling algorithm, namely Latent Dirichlet Allocation (LDA) \cite{Silva2021TopicResearch} to explore architectural topics based on their contained AK concepts. Using two research methods (i.e. grounded theory and LDA) to identify topics of architectural blogs supports exploring topics from different perspectives, as well as data triangulation.  Finally, we applied statistical analysis to identify significant co-occurrences between types and topics of architectural blogs, and the relevance of topics for certain ADD steps. In summary, we achieved the following contributions:
\begin{itemize}
\item The \textit{types of blogs} that contain AK as retrieved by the most popular Web search engine Google.
\item The \textit{topics of blogs}, their contained AK concepts, and their significant co-occurrences with blog types.
\item The \textit{relevance} of architectural blog articles from certain topics in following ADD steps.
\item A \textit{dataset} of 718 classified architectural blog articles, based on their types and topics.
\item A semi-automated \textit{approach to identify architectural topics} using LDA and AK ontology.
\end{itemize}
Section \ref{sec:background} provides a background on ADD. Section \ref{sec:studydesign} explains our study design, Sections \ref{sec:RQ1Results}, \ref{sec:RQ2Results}, and \ref{sec:RQ3Results} present our results. Sections \ref{sec:discussion}, \ref{sec:threats}, \ref{sec:relatedwork} discuss our results, threats to validity, and related work. Finally, Section \ref{sec:conclusion} concludes the paper.

\section{Background on the Attribute Driven Design}
\label{sec:background}
In this paper, we evaluate the relevance of architectural blogs to follow the ADD steps \cite{KazmanDesigningSoftware2016} (see Section \ref{sec:RQ3Results}). We have selected the ADD process, as it is the most well-established architecture design process, it provides concrete steps, and it was previously applied by Soliman et al. \cite{Soliman2021ExploringKnowledge}. %Kazman et al. \cite{KazmanDesigningSoftware2016} presented a seven-step, iterative process to design a software architecture. 
In this section, we provide a brief summary of three out of the seven ADD steps, namely those that involve making design decisions, and thus require AK:

%were used in Soliman et al. \cite{Soliman2021ExploringKnowledge}, in which participants executed queries in Google, searching for architectural information to solve tasks which performed one of these steps. The method specifies that multiple alternative design concepts need to be gathered and compared in order for a single solution to be selected and customised for a specific design issue. The explanation of each step is as follows:

% original: \textit{Identify design concepts}: In this step, alternative architectural solutions are identified for a design issue. For example, a software engineer might look for alternative broker technologies, which might fulfill system requirements and align with the system constraints. As an example, these alternative solutions for broker technologies could be RabbitMQ, Kafka and ActiveMQ. To perform this step, software engineers need to search for information regarding alternative solution options.

%paraphrased:
%space
\textit{Identify design concepts}: A list of alternative architectural solutions is identified. For example, a software engineer might identify a list of technology solutions (eg. Spark, Storm, Flink, etc.) when designing a stream processing engine.
%to ensure their choice aligns with their system requirements.

\textit{Select design concepts}: One design concept from the list of alternative solutions is selected. Each solution is compared and evaluated against requirements, constraints, quality attributes (e.g. security, performance, etc.), and considering the benefits and drawbacks of solutions. For example, a software engineer might require high performance of stream processing, she would therefore rather select Spark or Flink over Storm. 

% For example, a software engineer might search for the benefits and drawbacks of each solution, or even make use of analysis methods (eg. cost benefit analysis), to find the most appropriate option.

\textit{Instantiate architecture elements}: Elements of the selected architectural solution are modified to achieve system requirements. For example, a software engineer might configure a reference architecture to meet a quality attribute such as add a custom layer, or specify relationships and responsibilities of elements of an architectural pattern \cite{KazmanDesigningSoftware2016}. 

%(eg. defining a number of threads in a thread pool)

%%%%%%%%%%%%%%%%%%%%%%%%%%%%%%%%%%

\section{Study design}
\label{sec:studydesign}
\subsection{Research questions}
To achieve our aim (see Section \ref{sec:introduction}), we ask the following research questions (RQs):

\noindent\textit{(\textbf{RQ1}) What types of architectural blogs do Web search engines retrieve?}

Software engineers write architectural articles in blogs (e.g. personal blogs) with different characteristics. For instance, some blogs are hosted by software companies, and focus on specific solutions (e.g. Big Data technologies), while other blogs are hosted by software development communities and discuss multiple topics on different solutions. We ask this RQ to determine types of blogs where AK is shared, and their characteristics as retrieved by Web search engines (e.g. Google). By answering this RQ, researchers can develop approaches that classify and index architectural blogs to find and re-use AK. Moreover, practitioners can make informed decisions on the types of blogs to share and search for AK.

%\noindent\textit{(\textbf{RQ2}) What kind of topics are discussed in architectural blogs, and what AK concepts do they involve? and in what types of blogs are the topics being discussed?}
\noindent\textit{(\textbf{RQ2}) What kind of topics are discussed in the different types of architectural blogs?
%, and in what types of blogs are these topics discussed?
}

Blog articles discuss different topics (e.g. comparing solutions). Each topic involves different AK concepts.
Thus, we ask this RQ to determine topics discussed in architectural blogs, their involved AK concepts, and the types of blogs (e.g. personal blog) that discuss each topic. These topics can guide practitioners to search for and re-use certain AK concepts. They can also help researchers to develop approaches that automatically classify architectural blogs based on their topics.

%\noindent\textit{(\textbf{RQ3}) What topics of architectural blogs are retrieved by Web search engines to support the ADD steps, and how relevant are they to practitioners?}
\noindent\textit{(\textbf{RQ3}) How relevant are the topics of architectural blogs as retrieved by Web search engines to support practitioners follow the ADD steps?}

\begin{figure}
	\centering
        \includegraphics[scale=0.42]{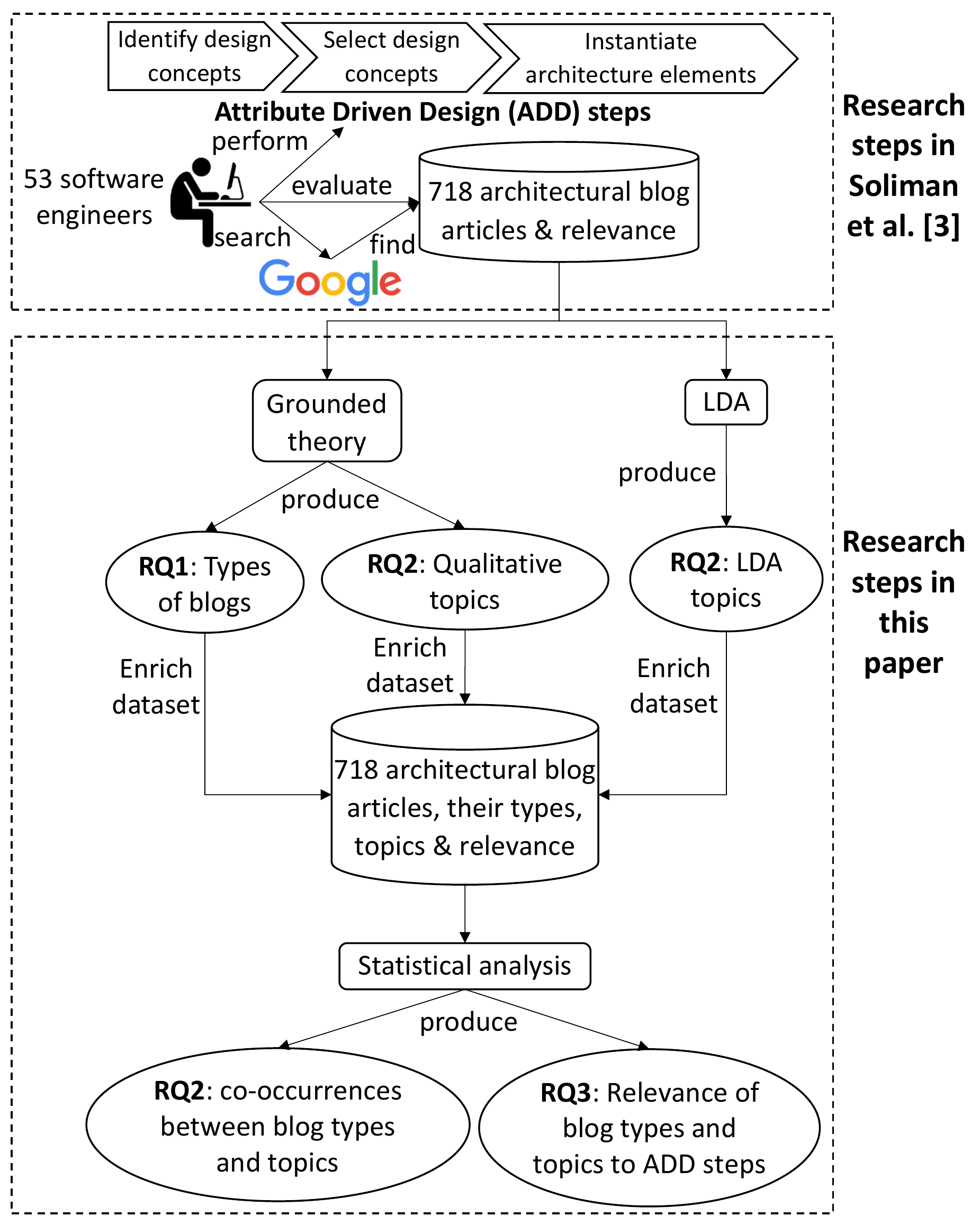}
	\caption{High level view on the research process to answer RQs}
	\label{fig:process}
\vspace{-5mm}
\end{figure}

Some topics of architectural blogs could be more useful for practitioners to conduct certain design steps (such as the ADD steps in Section \ref{sec:background}). However, Web search engines (e.g. Google) retrieve blog articles based on keyword-searches, and thus might not retrieve relevant results for each design step \cite{Soliman2021ExploringKnowledge}.
%because some topics of architectural blogs could be more relevant for practitioners than others. 
Therefore, we ask this RQ to determine topics of architectural blogs that Web search engines do retrieve, as well as which topics are relevant for conducting the ADD steps, according to practitioners. By answering this RQ, practitioners could target their search towards more relevant architectural blogs to achieve their design steps effectively. Moreover, researchers can adapt their AK finding approaches by promoting architectural blogs with the highest relevance. Figure \ref{fig:process} shows an overview on the research process to answer the RQs.
%Some topics could be useful for practitioners to conduct certain design steps.

%\noindent\textit{(\textbf{RQ4}) What AK concepts are shared in architectural blogs?}

%AK concepts constitute the building blocks of AK, which software engineers need to make ADDs. Thus, we ask this RQ to accurately identify the AK concepts, which software engineers share in architectural blogs. Moreover, the results of this RQ can benefit researchers to extract AK concepts from architectural blogs to automatically populate AK repositories.

%\subsection{Overview on the research process}
 %First, we applied grounded theory (see Section \ref{sec:RQ1Steps}) to identify types of blogs (answers RQ1), and qualitative topics of architectural blogs (answers RQ2). In parallel, we applied LDA (see Section \ref{sec:RQ2Steps}) to identify LDA topics of architectural blogs (answers RQ2). We note that we answer RQ2 using both qualitative (using grounded theory) and quantitative (using LDA) approaches. After these two steps, we conducted statistical analysis (see Section \ref{sec:RQ3Steps}) to determine co-occurrences between types and topics (answers RQ2), as well as relevance of architectural blog types and topics to the ADD steps (answers RQ3).
\subsection{Dataset of architectural blog articles}
\label{sec:dataset}
\begin{table}[]
\centering
\caption{Architectural tasks}
\label{tab:tasks}
\begin{tabular}{p{17mm} p{63mm}}
\hline
\textbf{ADD step} & \multicolumn{1}{c}{\textbf{Task description}} \\ 
\hline

Identify design concepts
          & $\bullet$ For a realtime dashboard, identify middleware technologies which scale to \textgreater 100k users \\ 

%\cline{2-3} 

          & 
% A system needs to communicate with mobile apps. Identify JSON parsers for Java with high performance, considering license constraints.
$\bullet$ Identify high performance Java JSON parsers for mobile app communication.
 \\ 
 
 \hline
 
Select design concepts                     & 
%A system communicates with a knowledge base via publish/subscribe patterns. Compare interoperability and latency of RabbitMQ, Kafka, and ActiveMQ.
$\bullet$ Compare interoperability and latency of RabbitMQ, Kafka, and ActiveMQ.
\\ 

%\cline{2-3} 

         & 
%Compare three technology families for big data systems: data collector, message brokers, and ETL engines. Requirements are throughput of 15,000 events/sec and availability of 99.99\%. 
$\bullet$ Compare data collectors, message brokers, and ETL engines for big data systems.
\\ 
\hline

\multirow{2}{17mm}{Instantiate \newline architecture \newline elements}
           & 
%CRM apps communicate with other systems using Apache Camel and RabbitMQ. Search for technology features and component designs to determine mechanisms for channeling, translation and routing, and deployment topology. 
$\bullet$ Search for technology features and component design to select deployment topology and routing.
%, translation and routing mechanism.
%Search for features to select a deployment topology and a translation and routing mechanism. 
\\ 
%\cline{2-3} 

          & 
%An application exposes services to other apps. Search for best practices regarding service decomposition to achieve high cohesion and low coupling. 
$\bullet$ Search for best practices regarding service decomposition to achieve high cohesion and low coupling.\\ 
\hline
\end{tabular}
\vspace{-4mm}
\end{table}
We re-used the dataset of architectural blog articles from the study of Soliman et al. \cite{Soliman2021ExploringKnowledge}. Following, we provide a brief description of that study, while further details are in \cite{Soliman2021ExploringKnowledge}. 
53 software engineers used Google searches to perform six architectural tasks (see Table \ref{tab:tasks}) that correspond to the three ADD steps (see Section \ref{sec:background}). A complete description of each task is available online \cite{OnlineMaterial}. To perform the six tasks, software engineers evaluated each web-page according to its relevance to an architectural task in a five-level Likert scale:
%We provide below brief definitions for the degrees of relevance. Further details on the study can be found in Soliman et al. \cite{Soliman2021ExploringKnowledge}:

\noindent $\bullet$ \textit{Very High Relevance (5)}: fulfills more than one requirement.

\noindent $\bullet$ \textit{High Relevance (4)}: fulfills one requirement of the task.

\noindent $\bullet$ \textit{Medium Relevance (3)}: provides information to the task.

\noindent $\bullet$ \textit{Low Relevance (2)}: is remotely relevant to the task.

\noindent $\bullet$ \textit{No Relevance (1)}: has no relevance information.

The 53 software engineers in Soliman et al. \cite{Soliman2021ExploringKnowledge} evaluated 2623 unique web-pages according to their relevance. Subsequently, 945 web-pages were classified as blogs and tutorials. These 945 web-pages provide the base dataset of this study.
%In order to prepare our dataset of architectural blogs for this study, 
We first verified whether these 945 web-pages are indeed architectural blogs, and excluded the following web-pages:

\noindent $\bullet$ Web-pages with \textit{no relevance (level 1 above)} to an architectural task as specified by practitioners. This is to ensure that all blog articles in the dataset are actually architectural.

\noindent $\bullet$ Blog articles in languages other than English.

\noindent $\bullet$ Web-pages that cannot be accessed anymore.

\noindent $\bullet$ Web-pages that include tutorials rather than blogs.

\noindent $\bullet$ Duplicated blog articles with different URLs.

As a result of this filtering step, a total of 718 unique architectural blog articles were used to answer the RQs.
\subsection{Apply grounded theory to identify types and topics of architectural blogs}
\label{sec:RQ1Steps}
To identify types and topics of architectural blogs, we manually analyzed all 718 of the architectural blog articles by following steps and best practices from grounded theory \cite{Strauss1990BasicsTechniques.,Stol2016GroundedGuidelines}. In detail, we performed the following three steps:
\subsubsection{Open coding}
\label{sec:RQ1OpenCoding}
To identify initial types and topics of blogs, the first and the second authors randomly selected a significant sample of 300 architectural blog articles from the 718 articles, and independently inspected each blog article as well as the web-sites, in which the blog articles are hosted. Both researchers inspected the 300 architectural blog articles through 3 iterations. Within each iteration, both researchers independently assigned a type and a topic for each blog article, by focusing on the content of the blog articles and avoiding being biased from literature;  
%Both researchers assigned one type of blogs to each web-page, 
subsequently they \textit{wrote memos} to explain why a type or a topic of blog has been determined for a certain article. For example, if all articles in a web-site are written by the same software engineer, then it is a personal blog, and if an article provides benefits and drawbacks of alternative solutions, then it is about comparing solutions. We share the memos for all blog articles online \cite{OnlineMaterial}. After each iteration, both researchers discussed the assigned types and topics of architectural blog articles, and performed \textit{constant comparisons} to determine differences between the different types and topics of blogs. %Accordingly, we determined attributes that differentiate types and topic of blogs. 
For instance, we determined the following attributes to differentiate the types of blogs:

\noindent $\bullet$ \textit{Number of authors who write blog articles in a web-site}. For example, personal blogs tend to have a single author.

\noindent $\bullet$ \textit{Host of the web-site}. For example, a blog article could be hosted by technology vendors or IT service companies.

\noindent $\bullet$ \textit{Relationship between authors of articles and the hosting web-site}. For example, authors of articles could be independent or employees of the company that hosts a web-site.

For the topics, we determined the following attributes:

\noindent $\bullet$ \textit{The purpose of an article}. Some articles discuss steps to design a system, while others provide steps to implement.

\noindent $\bullet$ \textit{The shared AK concepts in an article}. Some articles provide benefits and drawbacks of solutions, while other articles provide a list of alternative solutions to solve a problem.

\noindent $\bullet$ \textit{The number of discussed architectural solutions in an article}. Some articles discuss and compare multiple solutions, while other articles elaborate and evaluate a single solution.

As a result of this step, we defined initial types and topics of architectural blogs. During the third iteration, neither of the two researchers could determine new types or topics; consequently, this indicated \textit{theoretical saturation}.
\subsubsection{Axial coding}
\label{sec:RQ1SelectiveCoding}
Based on the initial types and topics from the open coding step, the first two authors of this paper classified the rest of the 718 architectural blog articles: the first author identified the topics for the rest of the 718 articles, while the second author identified the types for the rest of the 718 articles. For some uncertain cases of articles, both researchers discussed the article to agree upon a single type or topic. During discussions, both researchers noticed that some types of blogs caused disagreements between them. Thus, we merged these types of blogs to ensure good agreement between researchers. For example, some community blogs have sub-types (e.g. general versus technology-specific) which could be confusing to differentiate; therefore, we merged community blogs in a single type. By the end of this step, we reached the final set of \textit{types} and \textit{qualitative topics} of architectural blogs (see Sections \ref{sec:RQ1Results} and \ref{sec:RQ2ResultsQualitative}).
\subsubsection{Measure agreement}
\label{sec:RQ1Agreement}
To ensure the reliability of our qualitative analysis, we conducted a final agreement test: we randomly selected per author, a significant sample of 300 blog articles, which were not previously classified by that author. The first author independently classified the test sample regarding blog types, while the second author classified the test sample regarding topics. Based on this test sample, we calculated the kappa coefficient among the two authors, which are 0.81 regarding its type and 0.71 regarding the topics. These kappa values indicate good agreement beyond chance.
\subsection{Apply LDA to identify topics of architectural blogs}
\label{sec:RQ2Steps}
In the previous section, we applied grounded theory to identify qualitative topics of architectural blogs. In this Section, we applied LDA to identify topics (i.e. \textit{LDA topics}) in architectural blogs. Using LDA, we consider frequencies of terms and AK concepts within an article; these provide a different perspective on topics than qualitative analysis. We also note that we determined co-occurrences between the qualitative topics and LDA topics in Section \ref{sec:RQ3Steps}. For the LDA analysis, we followed steps and best practices for applying LDA in software engineering \cite{Silva2021TopicResearch}.
%As mentioned in Sect Nevertheless, architectural blog articles are involved and contain different AK concepts, which make them challenging and time consuming for researchers to manually identify all AK concepts to determine a topic. Therefore, to consider the whole content of blog articles,
%The advantage of our LDA approach that researchers can quickly replicate on a dataset of architectural blog articles.
In summary, we applied 5 steps. We performed the first three steps iteratively; subsequently, we performed steps 4 and 5.
%We applied the LDA algorithm in multiple iterations (among Steps 1-3): The iterations are only among steps first 3 steps. By the end of the iterations, we applied steps 4 and 5. In each iteration (among the first 3 steps), we updated the pre-processing step based on the results of the previous iteration, to refine and concretize architectural topics.
These five steps are elaborated in the following.
\subsubsection{Pre-process blogs}
\label{sec:RQ2preprocessBlogs}
Blog articles contain textual contents, which can mislead the LDA algorithm, resulting in identifying irrelevant and inconsistent topics. Thus, this step aims to remove and process textual contents in blog articles to guide LDA towards architecturally-relevant topics. 
We removed the following terms from blog web-pages:

\noindent $\bullet$ HTML tags using the Beautiful Soup library\footnote{\url{https://www.crummy.com/software/BeautifulSoup}}, stop words using the NLTK library\footnote{\url{https://www.nltk.org}}, and special characters (e.g. \#,-).

\noindent $\bullet$ Too frequent words that appear in more than 95\% of the blogs, and less frequent words that appear in less than 5\% of the blogs, because these terms bias LDA to form separate topics. Other researchers have also followed this step \cite{Silva2021TopicResearch}.

\noindent $\bullet$ Generic keywords that do not refer to an AK concept but could mislead the LDA algorithm in creating separate topics for them, such as ``microsoft", ``medium" and ``software". We expanded the list of generic keywords iteratively (see Step 3).

\noindent $\bullet$ Common terms related to source code implementation such as ``method", ``procedure", and ``array". The list of source code terms has been expanded iteratively (see Step 3). We removed source code terms, because we found that they bias LDA to create dedicated topics for blogs that contain source code elements; such topics are irrelevant to the actually discussed architectural topic.

After filtering out non-useful terms, we lematized terms to a single form. For example, ``deciding" and ``decided" are reduced to a single form ``decide". We used NLTK to lematize English terms. Moreover, we extended the NLTK dictionary to lematize architectural terms, which cannot be lematized by the default NLTK wordnet dictionary (e.g. ``scalability" and ``scalable" are reduced to a single form ``scale").

Subsequently, we found out that LDA was biased to identify topics based on their domain and technologies (e.g., Big Data and Middleware topics). Thus, to identify architecturally-relevant topics, independent of their domain, we replaced certain terms with tags that corresponds to their AK concepts. We considered specifically AK concepts from the ontology of AK in Stack Overflow \cite{Soliman2017DevelopingCommunities}. We have decided on this ontology due to the possible similarity of AK between Stack Overflow and blogs.
%Some terms refer to certain AK concepts. For example, RabbitMQ and Kafka are both technology solutions, while performance and security are both quality attributes. However, the LDA algorithm cannot recognize semantic relations between terms, and tends to be biased towards technological terms, rather than AK concepts. Thus, to support LDA algorithm identify topics in architectural blogs based on their discussed AK concepts, we replaced each term with a tag that refer to its respective AK concept. We considered specifically AK concepts from the ontology of AK in Stack Overflow \cite{Soliman2017DevelopingCommunities}. We have decided on this ontology due to the possible similarity between Stack Overflow and blogs. 
Table \ref{tab:AKConcepts_Examples} presents the list of selected AK concepts from the AK ontology \cite{Soliman2017DevelopingCommunities}, and examples of terms associated with each AK concept.
In addition to replacing terms with AK concepts, we associated the frequency of unique terms in a blog article with their respective tag of AK concept. For example, if a blog post discusses three technologies: Kafka, RabbitMQ and ActiveMQ, these will be replaced with $<$Technology\_Solution\_1$>$, $<$Technology\_Solution\_2$>$, and $<$Technology\_Solution\_3$>$. In this way, LDA can differentiate between blog articles that contain different terms from specific AK concepts. For example, blog articles which compare different technology solutions would have $<$Technology\_Solution$>$ tags with higher frequency, while blog articles that discuss different components will have $<$Component$>$ tag with higher frequency. We note that we extended the list of terms associated with each AK concept iteratively (see Step 3). We provide online \cite{OnlineMaterial} lists of terms that correspond to each AK concept, as well as scripts to replicate this pre-processing step.
\setlength{\tabcolsep}{2pt}
\begin{table}[]
\caption{Examples of terms for each AK concept}
\begin{tabular}{rl}
\hline
\multicolumn{1}{c}{\textbf{AK concept}}     & \textbf{Examples of terms}                                \\ \hline
\textless{}Technology\textgreater{}         & ActiveMQ, SQS, JSON, Kafka, Maven, Flume          \\ \hline
\textless{}Pattern\textgreater{}            & layer, publish subscribe, Microservice, ESB, SOA     \\ \hline
\textless{}Quality\_attribute\textgreater{} & performance, accuracy, usability, security, throughput    \\ \hline
\textless{}Requirement\textgreater{}        & business, dashboard, chart, customer, market, financial   \\ \hline
\textless{}Component\textgreater{}          & application, server, client, back end, front end \\ \hline
\textless{}Connector\textgreater{}          & retrieve, connect, consume, call, save, route, depend     \\ \hline
\textless{}Connector\_data\textgreater{}    & socket, payload, message, dump, token, call, update       \\ \hline
\end{tabular}
\label{tab:AKConcepts_Examples}
\vspace{-5mm}
\end{table}
\subsubsection{Apply LDA}
\label{sec:RQ2applyLDA}
After pre-processing blog articles, we experimented with a different number of topics ($k$), between 5 and 20 topics, and applied the most common LDA parameters ($\alpha=50/k$ and $\beta=0.01$) \cite{Silva2021TopicResearch}. We note that we further tuned LDA parameters in Step 4. After executing LDA on our dataset, each blog article is assigned to a single topic (with the highest coherence of terms and AK concepts). However, LDA does not describe or define each topic. Therefore, to understand each topic, we identified the top most frequently occurring terms and AK concepts in each topic. We used this list of terms and AK concepts in Step 3 to enrich the pre-processing step. Moreover, we used this list of terms and AK concepts in Step 5 to define \textit{LDA topics} of architectural blogs.
\subsubsection{Update pre-processing}
\label{sec:RQ2UpdatePreProcessing}
%We conducted the previous two steps (i.e. pre-process blogs and apply LDA) iteratively. For each iteration,
In this step, we checked the list of terms from the LDA topics (see Step 2), and determined general terms and source code terms that could be removed. Moreover, we determined new architectural terms that refer to certain AK concepts, which could be replaced with their respective AK concepts. Accordingly, we updated the pre-process blogs step with new terms to refine the LDA topics, and repeated the first three steps (i.e. pre-process blogs, apply LDA, and update pre-processing) to ensure that we identified all frequent terms from blog articles. We stopped repeating the three steps, once no new terms appeared in the list of the most occurring terms of each topic, which needed to be either removed or replaced with an AK concept.
\subsubsection{Tune LDA parameters}
\label{sec:RQ2TuneLDA}
After establishing the pre-processing of blogs, we experimented with a different number of topics up to 20 topics, and calculated the coherence. We noticed that we got higher coherence for number of topics between three and six topics, while the coherence decreased for more than six topics. Therefore, we decided to explore six topics of architectural blogs. Moreover, we experimented with $\alpha=50/k$ to $\alpha=1/k$, and calculated the density of topics among architectural blogs. Our experiments showed that the standard parameter of $\alpha=50/k$ provides a balanced density of topics among architectural blog articles.
\subsubsection{Define LDA topics}
\label{sec:RQ2LDATopics}
LDA represents topics as a set of occurring terms and AK concepts. Using these terms and AK concepts, we provided a definition of each LDA topic (see Section \ref{sec:RQ2ResultsLDA}). %researchers could determine main characteristics of the discussed topics \cite{Silva2021TopicResearch}. Accordingly, we provide a definition of each LDA topic based on the list of terms and AK concepts (see Section \ref{sec:RQ2ResultsLDA}).
\subsection{Apply statistical analysis to answer RQ2 and RQ3}
\label{sec:RQ3Steps}
\begin{figure*}
	\centering
        \includegraphics[scale=0.5]{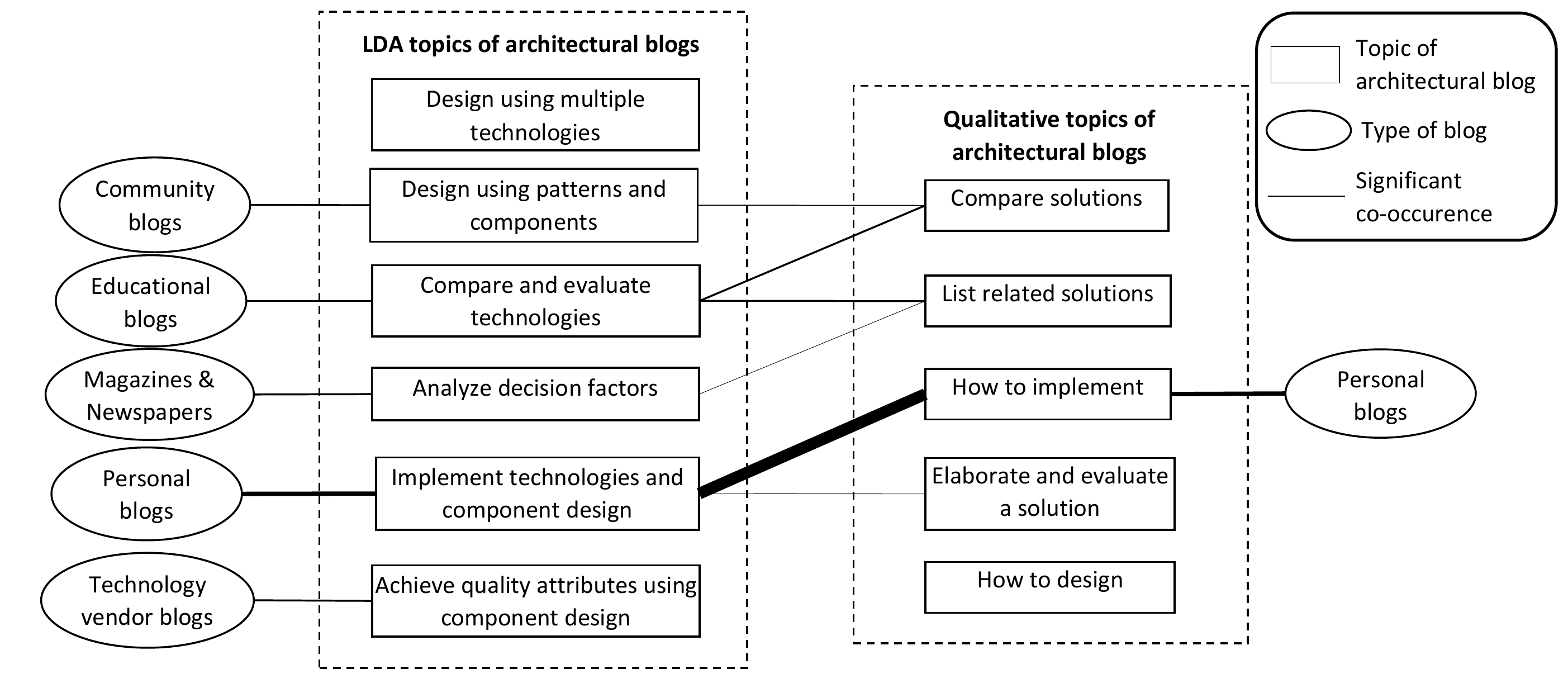}
	\caption{Significant co-occurrences between LDA topics and qualitative topics of architectural blogs, and between topics and types of blogs. The line thickness corresponds to the significance of co-occurrence as measured by the $\tilde{\chi}^2$ significance test.}
	\label{fig:Topics-co}
\vspace{-5mm}
\end{figure*}

As part of answering RQ2, we determine significant co-occurrences between qualitative topics, LDA topics and types of architectural blogs using the $\tilde{\chi}^2$ significant test \cite{PearsonSquare}. For each co-occurrence between a type or a topic, we calculated a 2x2 contingency table. For example between one type of blogs (e.g. personal blogs), and one topic of blogs (e.g. compare solutions), we calculate: number of co-occurring articles between the specified type and topic, number of articles with the specified type but with other topics, number of articles with the specified topic but with other types, number of articles with different types and topics. Among this 2x2 contingency table, we calculate the $\tilde{\chi}^2$ value, and determine significant co-occurrences with $\tilde{\chi}^2 > 10$ at p-value \textless 0.05. We present significant co-occurrences in Section \ref{sec:RQ2ResultsCooccurence} and Figure \ref{fig:Topics-co}.

To answer RQ3, we used the relevance indicated by the 53 software engineers in Soliman et al. \cite{Soliman2021ExploringKnowledge} to determine relevant articles for each ADD step. Specifically, we used descriptive statistics to calculate the number of relevant blog articles and their relevance for each topic when following the ADD steps. Moreover, we conducted a Kruskal-Wallis H test \cite{Kruskal1952UseAnalysis} to determine significant relevance between the different topics. We decided on the Kruskal-Wallis H test over the Anova test \cite{Fisher1992StatisticalWorkers}, because it can test abnormally distributed data as in our dataset. Further details about the significance test are shared online \cite{OnlineMaterial}.
We present the results in Section \ref{sec:RQ3Results}.
\vspace{-1mm}
\section {\textbf{RQ1}: Types of Architectural Blogs}
\label{sec:RQ1Results}
We analyzed the architectural blogs in our dataset using grounded theory (see Section \ref{sec:RQ1Steps}), and identified the following types of architectural blogs (in parentheses their respective percentages in the dataset), as retrieved by Google:

\noindent $\bullet$ \textbf{Community blogs (43\%)}: hosted by open publishing platforms (e.g. \textit{\url{dzone.com}}), and authored by many different software engineers, who volunteer to share their experience in blog articles. The majority of community blogs host a wide range of architectural topics such as those in \textit{\url{dzone.com}} and \textit{\url{medium.com}}. But some community blogs focus on specific topics such as data integration as in \textit{\url{dataintegrationinfo.com}}.

\noindent $\bullet$ \textbf{Technology vendor blogs (25\%)}: hosted by specific technology vendors, such as commercial companies (e.g. SAP) or open source foundations (e.g. Apache). Blog articles focus on topics related to the products or technologies produced by the hosting technology vendor such as \textit{\url{blogs.sap.com}} and \textit{\url{blog.rabbitmq.com}}. The authors of articles are experts and developers of these technologies.

\noindent $\bullet$ \textbf{Personal blogs (15\%)}: hosted and authored by individual software experts, and reflecting their personal experience such as articles in \textit{\url{martinfowler.com}} and \textit{\url{alexanderdevelopment.net}}.

\noindent $\bullet$ \textbf{IT service blogs (11\%)}: hosted by IT service companies that provide services such as software development or consulting (e.g. \textit{\url{openlogic.com/blog}}). Articles are posted by employees of these companies.

\noindent $\bullet$ \textbf{Magazines and newspapers blogs (3\%)}: hosted by specialized magazines or newspapers (e.g. \textit{\url{opensourceforu.com}}), where specific authors, possibly hired, write articles.

\noindent $\bullet$ \textbf{Educational blogs (3\%)}: hosted by training organizations or universities such as \textit{\url{edureka.co/blog}}. Articles are posted by students or educators.
\vspace{-2mm}
\begin{tcolorbox}[boxsep=1pt,left=10pt,right=10pt,top=3pt,bottom=3pt]
\textbf{RQ1 key takeaways}: \\
$\bullet$ Software engineers share their AK in different types of blogs with different hosting and authorship policies. \\
$\bullet$ Google search results contain architectural articles mostly from community blogs, followed by technology vendor blogs, and then personal blogs.
\end{tcolorbox}
\vspace{-3mm}
\section {\textbf{RQ2}: Topics of architectural blogs}
\label{sec:RQ2Results}
We analyzed architectural blog articles using grounded theory (see Section \ref{sec:RQ1Steps}) and LDA (see Section \ref{sec:RQ2Steps}). Each approach produced separate but related topics. Subsequently, we determined significant co-occurrences between the topics from the two approaches, as well as the types of blogs (see Section \ref{sec:RQ3Steps}). Figure \ref{fig:Topics-co} shows an overview on the qualitative and LDA topics, and their significant co-occurrences with each other and with blog types according to the $\tilde{\chi}^2$ significance test. 
In sub-sections \ref{sec:RQ2ResultsQualitative} and \ref{sec:RQ2ResultsLDA}, we respectively explain the qualitative and LDA topics, and their contained AK concepts. Moreover, we support our explanations with examples of articles (see Section ``References of blog articles"), referenced as [B\#] (e.g. \citeB{Elaborateandevaluateasolution1}). Finally, we discuss the co-occurrences between the topics and types of blogs in sub-section \ref{sec:RQ2ResultsCooccurence}.
\vspace{-2mm}
\subsection {Qualitative topics of architectural blogs}
\label{sec:RQ2ResultsQualitative}

We explain the qualitative topics (in parentheses their respective percentages in the dataset), as retrieved by Google (see Section \ref{sec:dataset}), as well as their contained \textit{AK concepts} (underlined in the text).
%the tag \textit{[AK concept]}.

\noindent $\bullet$ \textbf{Elaborate and evaluate a solution (28\%)}: Articles in this topic elaborate and evaluate a single \underline{architectural solution}, such as a technology and its features \citeB{Elaborateandevaluateasolution1}, or a pattern and its elements \citeB{Elaborateandevaluateasolution2}, or an architecture and its components \citeB{Elaborateandevaluateasolution4}, or a design principle and its application \citeB{Elaborateandevaluateasolution3}. For example, in \citeB{Elaborateandevaluateasolution1}, the author first elaborates the main \underline{features} of Amazon SQS technology such as ``\textit{SQS is a distributed, queue-based, messaging service...It supports building (loosely coupled) integrations between two applications}", then the author discussed abilities of Amazon SQS to achieve \underline{quality attributes} such as performance and availability. Another example is \citeB{Elaborateandevaluateasolution2}, where the author elaborates the Microservices architecture, and discusses its \underline{benefits and drawbacks}.

\noindent $\bullet$ \textbf{List of related solutions (20\%)}: Articles in this topic provide a list of related \underline{architectural solutions} such as lists of technologies \citeB{List2}, patterns \citeB{List3}, tactics \citeB{List6}, design principles \citeB{List4}, architectural components \citeB{List5}, and best practices \citeB{List1}. Solutions could be either alternative solutions to solve a \underline{design issue}, or solutions that complement each other to design an architecture of a system. Each solution could be associated with a description of its characteristics. In these articles, the solutions are not compared to each other. For example, \citeB{List1} provides a numbered list of 10 Microservices Best Practices, giving a short description of each item.

%\textit{e.g.2 Best 6 stock market APIs for 2022 ... Below are a list of good stock API that we have curated ...}\citeB{List2}

\noindent $\bullet$ \textbf{Compare solutions (20\%)}: Articles in this topic compare two or more \underline{architectural solutions} such as technologies \citeB{Comparesolutions1} or patterns \citeB{Comparesolutions2}, list their \underline{benefits and drawbacks}, and explain in which cases each could be applied. For example, \citeB{Comparesolutions2} defines two \underline{solutions}: Synchronous and Asynchronous request handling, states factors to decide on them (i.e. \underline{requirements and constraints}), and states their \underline{benefits and drawbacks}.
%\textit{``When should we use Async/Await method"} as well as the\textit{``Benefits -"} and \textit{``Limitations of async/await programming"}. 
Another example is \citeB{Comparesolutions1}, which compares several JSON libraries (i.e. \underline{architectural solutions}) based on their performance (i.e. \underline{quality attribute}).
%, and \textit{``according to the test results, we can select the most suitable JSON library according to the actual application scenario."}. 

\noindent $\bullet$ \textbf{How to design (18\%)}: Articles in this topic discuss different steps or \underline{issues} to design a system within a certain domain (e.g. Web applications \citeB{Howtodesign3}), or to design a system based on specific \underline{architectural solutions} such as technologies \citeB{Howtodesign2} or patterns \citeB{Howtodesign4}. The explanation of steps or issues are commonly supported with \underline{use-cases}, and alternatives of \underline{architectural solutions}. For example, \citeB{Howtodesign5} presents a \textit{``Seven Step API Design Methodology"}, within the domain of the Web, using the HTTP protocol. Another example is \citeB{Howtodesign6}, which explains how to design a routing topology using RabbitMQ, and discusses design issues, use cases and evaluations against quality attributes such as performance and scalability.

\noindent $\bullet$ \textbf{How to implement (14\%)}: Articles in this topic provide guidelines to implement a conceptual \underline{architectural solution} (e.g. component design \citeB{Howtoimplement3} or \underline{pattern} \citeB{Howtoimplement2}) through certain \underline{technologies}, or provide guidelines to integrate different technologies together \citeB{Howtoimplement1}. For example, \citeB{Howtoimplement2} explains how to implement a RESTful web service, specifically in Java using JSON-B and JSON-P. Another example is \citeB{Howtoimplement1}, which explains how to integrate Apache Camel with the CData JDBC Driver to copy Dynamics CRM data to a JSON file on disk.
\vspace{-2mm}
\subsection {LDA topics of architectural blogs}
\label{sec:RQ2ResultsLDA}
We define LDA topics based on the most commonly occurring terms and AK concepts in each topic, as identified by the LDA algorithm. Table \ref{tab:LDATopics} provides top terms, AK concepts and their frequencies (columns freq.) among articles for each LDA topic. Also, number of unique terms that belong to each AK concept per blog article (column terms/article).

\noindent $\bullet$ \textbf{Design using patterns and components (24\%)}: Articles in this topic contain terms that refer to designing an architecture, as well as names of architectural patterns, components and quality attributes. For example, \citeB{Designusingpatternsandcomponents1} discusses different patterns and configurations (e.g. number of layers) to design web applications. Another example is \citeB{Designusingpatternsandcomponents2}, which discusses over 12 design patterns to design microservice systems.

\noindent $\bullet$ \textbf{Achieve quality attributes using component design (22\%)}: Articles in this topic contain terms that refer to quality attributes, as well as terms that refer to components, connectors and patterns. For example \citeB{Achievequalityattributesusingcomponentsdesign1} explains how to achieve performance requirements through different component design using Apache Kafka. Another example is \citeB{Achievequalityattributesusingcomponentsdesign2}, which compares RabbitMQ and Apache Kafka regarding their abilities to realize a component design that achieves certain quality attributes (e.g. scalability and performance).

\noindent $\bullet$ \textbf{Implement technologies and component design (20\%)}: Articles in this topic contain names of technologies, and terms that refer to implementation details (e.g. ``example"), as well as terms that refer to components and connectors. For example, \citeB{Implementtechnologiesandcomponentsdesign1} explains implementation details for Microservice architecture using Spring technologies. Another example is \citeB{Implementtechnologiesandcomponentsdesign2}, which explains how to implement event-driven integration between components using Apache Camel and Kubernetes.

\noindent $\bullet$ \textbf{Analyze decision factors (16\%)}: Articles in this topic contain terms that refer to factors that influence design decisions. These involve business requirements, constraints (e.g. ``time"), and benefits of solutions (e.g. using terms ``free" and ``easy"). For example \citeB{Analyzedecisionfactors1} discusses loading data to the cloud
%, and whether this data should be loaded directly or after transformation
. Several decision factors are discussed such as productivity, infrastructure complexity and performance. Another example is \citeB{Analyzedecisionfactors2}, which discusses the integration between two technologies: SAP and power Apps, and explain the business benefits from such integration in terms of real time visibility and efficiency.

%as well as terms that refer to analysis and decision making such as ``analysis" and ``decision". 
%Moreover, articles in this topic contain terms about the benefits of solutions such as ``free", ``best", ``easy", and ``benefit".

\noindent $\bullet$ \textbf{Compare and evaluate technologies (13\%)}: Articles in this topic contain terms that refer to comparing technologies such as ``vs", ``feature", and names of different technologies, as well as terms that refer to quality attributes and requirements. For example, in \citeB{Compareandevaluatetechnologies1}, the two technologies ActiveMQ and RabbitMQ are compared regarding their features and quality attributes, such as scalability and interoperability.

\noindent $\bullet$ \textbf{Design using multiple technologies (5\%)}: Articles in this topic contain terms that refer to designing an architecture, as well as terms that refer to requirements and technologies. For example, \citeB{Designusingmultipletechnologies1} explains the architecture of web applications using Microsoft technologies.

% \usepackage{multirow}
\setlength{\tabcolsep}{2pt}
\begin{table}[]
\caption{Top terms, AK concepts, their frequencies (freq.), and number of unique terms per article (terms/article) for LDA topics.}
\begin{tabular}{|p{1.5cm} cC{0.6cm}|C{2.4cm}C{0.6cm}c|}
\hline
\textbf{}                                                           \textbf{LDA Topic}  & \textbf{Terms} & \textbf{freq.} & \textbf{AK concepts}                        & \textbf{freq.} & \textbf{terms/article} \\ \hline
\multirow{5}{1.5cm}{\textit{Design using patterns and components}}        & architecture   & 2253           & \textless{}Pattern\textgreater{}            & 4514           & 2 to 5                 \\
                                                                      & design         & 1617           & \textless{}Component\textgreater{}          & 4509           & 6 to 16                \\
                                                                      & need           & 1250           & \textless{}Quality\_attribute\textgreater{} & 1226           & 2 to 4                 \\
                                                                      & web            & 1103           & \textless{}Connector\_data\textgreater{}    & 1010           & 5 to 6                 \\
                                                                      & approach       & 814            & \textless{}Connector\textgreater{}          & 302            & 7                      \\ \hline
\multirow{5}{1.8cm}{\textit{Achieve quality attributes using component design}} & time           & 1213           & \textless{}Component\textgreater{}          & 7430           & 6 to 16                \\
                                                                      & exchange        & 1104           & \textless{}Connector\_data\textgreater{}    & 2031           & 4 to 10                \\
                                                                      & need           & 1021           & \textless{}Pattern\textgreater{}            & 2041           & 2 to 4                 \\
                                                                      & topic          & 1010           & \textless{}Connector\textgreater{}          & 907            & 7 to 10                \\
                                                                      & support        & 716            & \textless{}Quality\_attribute\textgreater{} & 805            & 2 to 3                 \\ \hline
\multirow{5}{1.5cm}{\textit{Implement technologies and component design}}       & create         & 1103           & \textless{}Technology\textgreater{}         & 7411           & 6 to 31                \\
                                                                      & example        & 1121           & \textless{}Connector\_data\textgreater{}    & 2212           & 4 to 7                 \\
                                                                      & version        & 801            & \textless{}Connector\textgreater{}          & 1607           & 6 to 13                \\
                                                                      & type           & 721            & \textless{}Component\textgreater{}          & 713            & 7 to 9                 \\
                                                                      & test           & 451            & \textless{}Requirement\textgreater{}        & 301            & 4                      \\ \hline
\multirow{5}{1.5cm}{\textit{Analyze decision factors}}                    & time           & 1132           & \textless{}Requirement\textgreater{}        & 8741           & 4 to 14                \\
                                                                      & real           & 901            & \textless{}Component\textgreater{}          & 2028           & 6 to 18                \\
                                                                      & report         & 851            & \textless{}Connector\_data\textgreater{}    & 402            & 4                      \\
                                                                      & analysis       & 801            &                                             &                &                        \\
                                                                      & free           & 703            &                                             &                &                        \\ \hline
\multirow{5}{1.5cm}{\textit{Compare and evaluate technologies}}           & vs             & 1803           & \textless{}Technology\textgreater{}         & 4721           & 6 to 23                \\
                                                                      & support        & 1103           & \textless{}Requirement\textgreater{}        & 1328           & 5 to 11                \\
                                                                      & feature        & 946            & \textless{}Component\textgreater{}          & 1316           & 6 to 13                \\
                                                                      & time           & 810            & \textless{}Quality\_attribute\textgreater{} & 953            & 2 to 4                 \\
                                                                      & offer          & 510            & \textless{}Pattern\textgreater{}            & 407            & 4                      \\ \hline
\multirow{5}{1.5cm}{\textit{Design using multiple technologies}}          & question       & 601            & \textless{}Technology\textgreater{}         & 1553           & 6 to 27                \\
                                                                      & developer      & 506            & \textless{}Requirement\textgreater{}        & 608            & 4 to 12                \\
                                                                      & development    & 450            & \textless{}Component\textgreater{}          & 206            & 10                     \\
                                                                      & design         & 406            &                                             &                &                        \\
                                                                      & architecture   & 403            &                                             &                &                        \\ \hline
\end{tabular}

\label{tab:LDATopics}
\vspace{-6mm}
\end{table}
\vspace{-2mm}
\subsection {Co-occurrences of types and topics of architectural blogs}
\label{sec:RQ2ResultsCooccurence}
Figure \ref{fig:Topics-co} shows significant co-occurrences between the LDA topics (see Section \ref{sec:RQ2ResultsLDA}), the qualitative topics (see Section \ref{sec:RQ2ResultsQualitative}), and the types of blogs (see Section \ref{sec:RQ1Results}). 
%Moreover, Figure \ref{fig:Topics-co} shows significant co-occurrences between the topics of architectural blogs and the types of blogs (see Section \ref{sec:RQ1Results}). 
From the co-occurrences, we can observe the following: 

\noindent $\bullet$ \textit{Some LDA topics logically co-occur with their corresponding qualitative topics}. For instance, the LDA topic \textit{Implement technologies and component design} significantly co-occurs with the \textit{How to implement} qualitative topic. Both topics focus on the implementation of an architecture. Similarly, the LDA topic \textit{Compare and evaluate technologies} significantly co-occurs with the \textit{Compare solutions} qualitative topic. Both topics focus on comparing solutions. This confirms that our proposed approach that uses LDA and AK ontology (see Section \ref{sec:RQ2Steps}) succeeded to identify some topics that correspond to their qualitative counterpart.

\noindent $\bullet$ \textit{Some LDA and qualitative topics complement each other}. On the one hand, LDA topics focus on certain types of architectural solutions (e.g. either technologies or patterns). On the other hand, qualitative topics focus on the purpose of an architectural blog. For example,  \textit{Design using patterns and components} significantly co-occurs with \textit{Compare solutions}. This indicates that software engineers discuss the comparison of patterns and components to design an architecture of a system in these articles. Another example is \textit{Implement technology and architecture}, which significantly co-occurs with \textit{Elaborate and evaluate solution}. This indicates that software engineers provide implementation details (e.g. source code) to elaborate a solution, and to evaluate a solution (e.g. provide source code to benchmark a technology).

\noindent $\bullet$ \textit{Most blog types significantly co-occur with a single topic}. With the exception of IT service blogs, four blog types significantly co-occur with one topic, and one blog type (\textit{Personal blogs}) co-occurs with two topics. For instance, articles in \textit{Community blogs} significantly co-occur with the \textit{Design using patterns and components} LDA topic. While articles in \textit{Personal blogs} significantly co-occur with the \textit{Implement technology and architecture topic} and \textit{How to implement} topics. This finding can help software engineers to find AK. For example, to find AK about patterns and components, software engineers should better search in community blogs, while to find AK related to the implementation of an architecture, software engineers should better search in personal blogs.

\noindent $\bullet$ \textit{Few LDA and qualitative topics do not significantly co-occur with other topics}. The LDA topics \textit{Achieve quality attributes using component design} and \textit{Design using multiple technologies}, as well as the qualitative topic \textit{How to design} do not significantly co-occur with other topics. Thus, articles in these topics either rarely co-occur with each other or equally co-occur with several other topics, but not one to a significant degree. For example, the \textit{How to design} qualitative topic involves long articles that discuss patterns, components, technologies, and decisions factors. These articles equally co-occur with the LDA topics, because they do not focus on a specific architectural solution.

\begin{tcolorbox}[boxsep=1pt,left=10pt,right=10pt,top=3pt,bottom=3pt]
\textbf{RQ2 key takeaways}: \\
$\bullet$ \textit{Listing, elaborating, evaluating, and comparing architectural solutions} present the majority of AK (68\%) in architectural blogs (see Section \ref{sec:RQ2ResultsQualitative}). \\
$\bullet$ \textit{Architectural patterns, component design and principles} are discussed in the majority (46\%) of architectural blogs, followed by discussions on \textit{technologies} and their implementations (38\%) (see Section \ref{sec:RQ2ResultsLDA}). \\
$\bullet$ \textit{Using LDA in combination with AK ontology} identifies reasonable architectural topics that correspond and complement the qualitative topics (see Section \ref{sec:RQ2ResultsCooccurence}). \\
$\bullet$ Each \textit{type of blog} is specialized in one or two topics.
\end{tcolorbox}
\section {\textbf{RQ3}: Relevance of Architectural Blog Topics to Attribute Driven Design Steps} 
\label{sec:RQ3Results}
In the study of Soliman et al. \cite{Soliman2021ExploringKnowledge}, 53 software engineers evaluated the relevance of blog articles to perform tasks that apply the ADD steps (see Section \ref{sec:dataset}). We use this data to determine what topics of architectural blogs (see Section \ref{sec:RQ2Results}) are retrieved by the Google search engine, and what topics are most relevant for software engineers to apply the ADD steps (see Section \ref{sec:RQ3Steps}). In the following sub-sections, we present the number of relevant blog articles (as retrieved by Google), and their relevance to each of the ADD steps, across the different topics (see Sections \ref{sec:RQ3ResultsQualitativeTopics} and \ref{sec:RQ3ResultsLDATopics}).
\vspace{-4mm}
\begin{figure}[ht]
\centering
\begin{subfigure}{0.5\textwidth}
  \centering
  % include first image
  \includegraphics[scale=0.46]{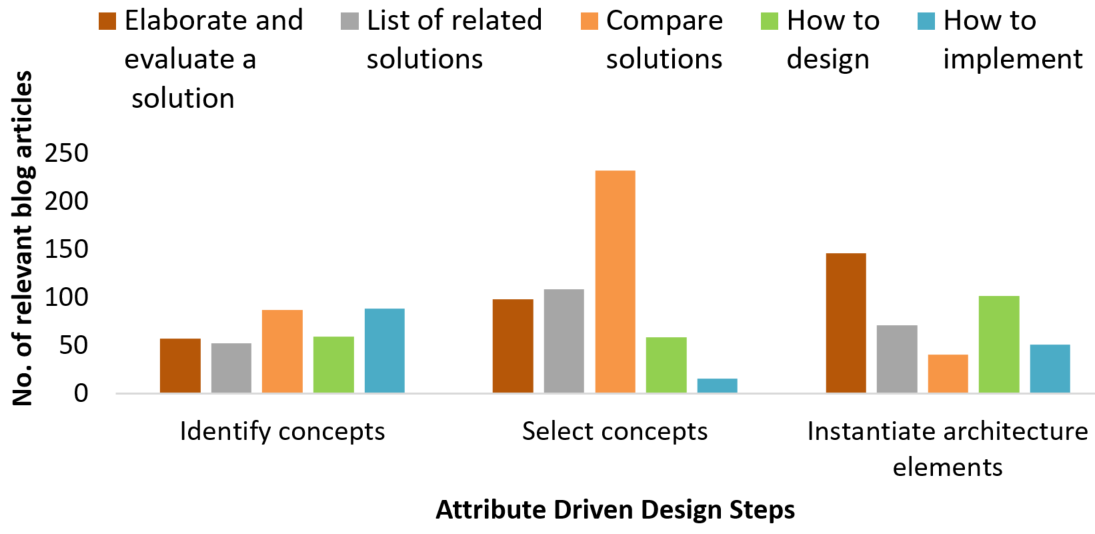}  
  \caption{Number (No.) of relevant blog articles per design step as retrieved by Google}
  \label{fig:qual_topic_relevance_count}
\end{subfigure}
\begin{subfigure}{0.5\textwidth}
  \centering
  % include second image
  \includegraphics[scale=0.75]{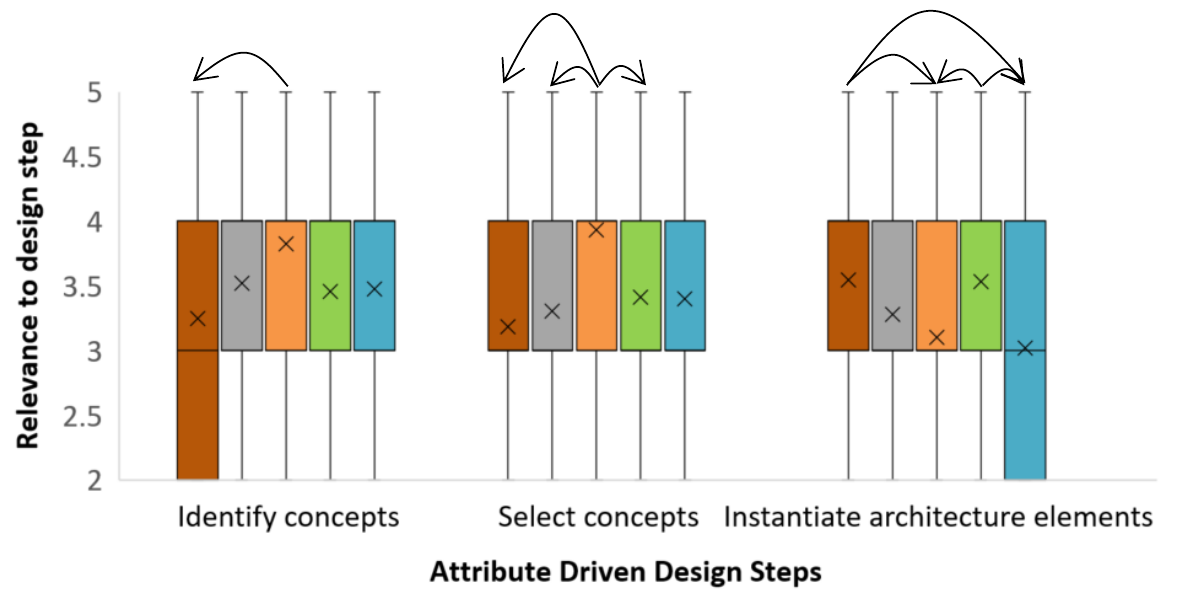}  
  \caption{Relevance of blog articles in each topic per ADD step, X presents the mean of relevance, and arrows refer to significant differences.}
  \label{fig:qual_topic_relevance_rel}
\end{subfigure}
\caption{Number and relevance of qualitative topics per ADD step}
\label{fig:qual_topic_relevance}
\vspace{-3mm}
\end{figure}
\vspace{-2mm}
\subsection {Relevance of qualitative topics}
\label{sec:RQ3ResultsQualitativeTopics}
Figure \ref{fig:qual_topic_relevance} shows the number of relevant blog articles for each qualitative topic as retrieved by Google, and their relevance as specified by practitioners. From Figure \ref{fig:qual_topic_relevance}, we can observe that Google search results are different among the three ADD steps, and practitioners evaluate articles from different topics differently for each ADD step:

\noindent $\bullet$ For \textit{Identify design concepts}, Google tends to retrieve the five topics of blogs similarly (see Figure \ref{fig:qual_topic_relevance_count}). However, our significance test shows that \textit{Compare solutions} is significantly more relevant than \textit{Elaborate and evaluate a solution} (see Figure \ref{fig:qual_topic_relevance_rel}). Thus, Google tends to retrieve blog articles with low relevance from the \textit{Elaborate and evaluate a solution} topic, which make it challenging for practitioners to find relevant AK for this step.

\noindent $\bullet$ For \textit{Select concepts}, Google tends to retrieve articles from the \textit{Compare solutions} topic significantly more than other topics (see Figure \ref{fig:qual_topic_relevance_count}). The Google search results are reasonable for this step, and align with our significance test, which shows that articles from the \textit{Compare solutions} topic are significantly more relevant to this step than other topics (see Figure \ref{fig:qual_topic_relevance_rel}).

\noindent $\bullet$ For \textit{Instantiate architecture element}, Google retrieves mostly articles from the \textit{Elaborate and evaluate a solution} topic, followed by the \textit{How to design} topic (see Figure \ref{fig:qual_topic_relevance_count}). Both topics have the highest relevance according to our significance test (see Figure \ref{fig:qual_topic_relevance_rel}). However, Google retrieves some blog articles of low relevance from the \textit{How to implement} and the \textit{Compare solutions} topics, which negatively affect the effectiveness of Google to find AK for this step.
%while promoting articles from the \textit{How to design} topic that showed significance.

%From Figure \ref{fig:qual_topic_relevance_count}, we can observe that the \textit{compare solutions} and the \textit{List of related solutions} topics have the most articles to support the select design concepts step, and the \textit{Elaborate and evaluate a solution} and the \textit{How to design} topics have the most articles to support the instantiate architecture elements step. Moreover, the \textit{How to implement} articles can mostly support the identify design concepts step. From Figure \ref{fig:qual_topic_relevance_rel}, we can observe that some qualitative topics have different mean of relevance to support the same design step. For instance, the \textit{Compare solutions} have higher mean of relevance compared to the \textit{List of related solutions} topic to make the select design concepts step. Moreover, the \textit{How to implement} topic has the lowest relevance compared to other topics to support the instantiate architecture elements step.
\vspace{-3mm}
\begin{figure}[ht]
\centering
\begin{subfigure}{0.5\textwidth}
  \centering
  % include first image
  \includegraphics[scale=0.51]{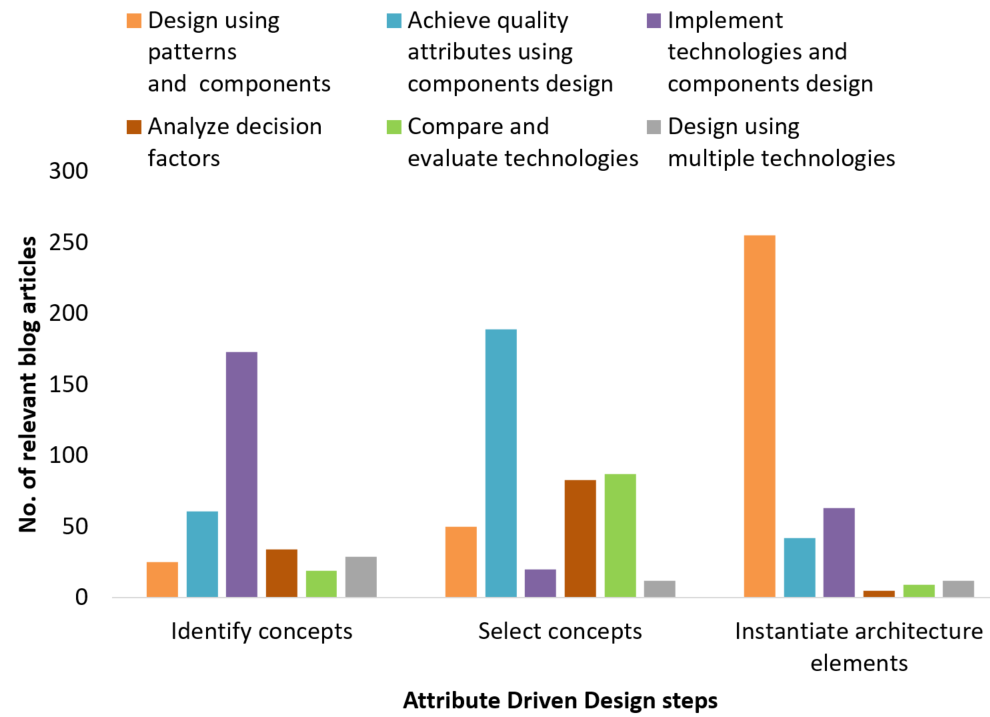}  
  \caption{Number (No.) of relevant blog articles per design step as retrieved by Google}
  \label{fig:LDA_topic_relevance_count}
\end{subfigure}
\begin{subfigure}{0.5\textwidth}
  \centering
  % include second image
  \includegraphics[scale=0.75]{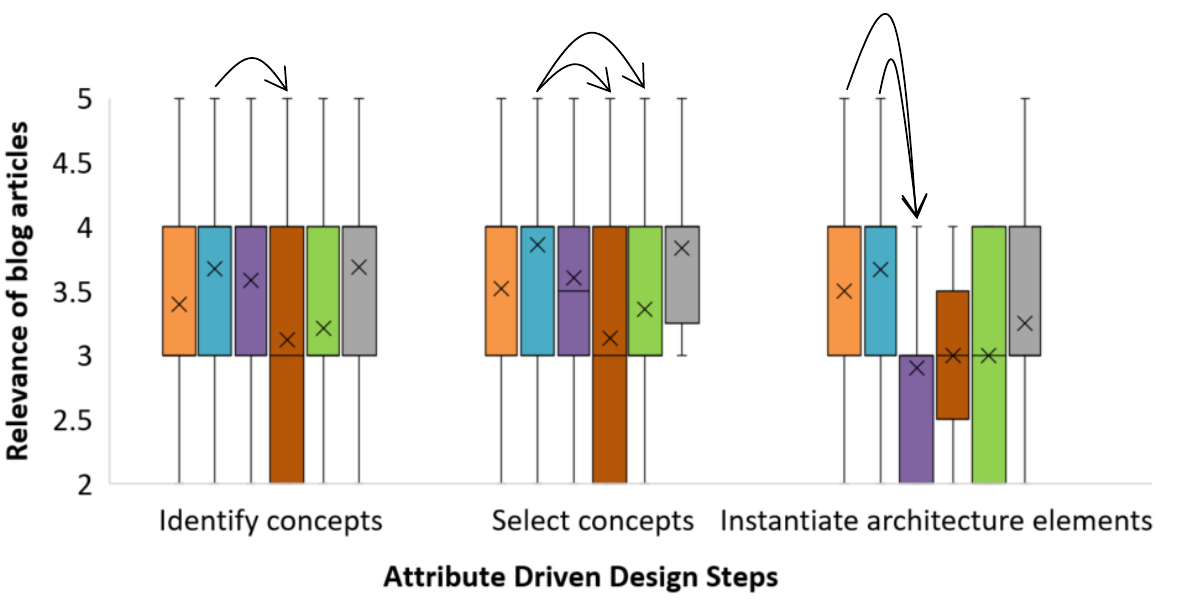}  
  \caption{Relevance of blog articles in each topic per ADD step, X presents the mean of relevance, and arrows refer to significant differences.}
  \label{fig:LDA_topic_relevance_rel}
\end{subfigure}
\caption{Number and relevance of architectural blog articles for each LDA topic, and for each Attribute Driven Design step}
\label{fig:LDA_topic_relevance}
\vspace{-3mm}
\end{figure}
\vspace{-2mm}
\subsection {Relevance of LDA topics}
\label{sec:RQ3ResultsLDATopics}
Figure \ref{fig:LDA_topic_relevance} shows the number of relevant blog articles for each LDA topic as retrieved by Google, and their relevance as specified by practitioners. From Figure \ref{fig:LDA_topic_relevance_count}, we can observe that Google tends to predominantly retrieve blog articles from a specific LDA topic to support an ADD step. Similarly to the qualitative topics, practitioners evaluate articles from different LDA topics differently for each ADD step (see Figure \ref{fig:LDA_topic_relevance_rel}):

\noindent $\bullet$ For \textit{Identify design concepts}, Google retrieves most blog articles from the \textit{Implement technologies and component design} topic. However, our significance test shows that the topic \textit{Achieve quality attributes using component design} is significantly relevant to this step as well. Moreover, some articles that belong to the \textit{Analyze decision factors} topic have significantly low relevance for this ADD step. Thus, search approaches should consider to filter out blog articles from the \textit{Analyze decision factors} topic to make this step effectively.

\noindent $\bullet$ For \textit{Select design concepts} Google retrieves the majority of blog articles from the \textit{Achieve quality attributes using component design} topic, followed by articles from the \textit{Analyze decision factors} and \textit{Compare and evaluate technologies}. However, our significance test shows that the topic \textit{Achieve quality attributes using component design} is significantly more relevant than both topics \textit{Analyze decision factors} and \textit{Compare and evaluate technologies}. Thus, search approaches can filter out irrelevant articles from these two topics to ensure effective search for this ADD step.

\noindent $\bullet$ For \textit{Instantiate architecture elements} Google tends to retrieve predominantly articles from the \textit{Design using patterns and components} topic. However, our significance test shows that articles from the \textit{Achieve quality attributes using component design} topic are also significant to this design step. In contrast, articles from \textit{Implement technologies and component design} have significantly low relevance to this step, which could be filtered out by approaches to ensure effective search for AK.
%For instance, the \textit{Implement technologies and connectors} LDA topic supports the identify design concept step, the \textit{Compare and evaluate technologies} LDA topic supports the select design concept step, and the \textit{Design using patterns and components} LDA topic supports the Instantiate architecture elements step. From Figure \ref{fig:LDA_topic_relevance_rel}, we can observe that articles that belong to certain LDA topics have different relevance to support the same design step. For example, the relevance of articles in the \textit{Design using patterns and components} LDA topic are higher than articles in the \textit{Implement technologies and connectors} LDA topic to support the instantiate architecture element step. On the other hand, the mean of relevance for articles in the \textit{Achieve constraints using component design} LDA topic is higher than the mean of relevance for articles in the \textit{Compare and evaluate technologies} LDA topic to support the select design concepts step.
\vspace{-1mm}
\begin{tcolorbox}[boxsep=1pt,left=5pt,right=5pt,top=3pt,bottom=3pt]
\textbf{RQ3 key takeaways}: \\
$\bullet$ Practitioners evaluate the relevance of topics differently for each design step (see Figures \ref{fig:qual_topic_relevance_rel} and \ref{fig:LDA_topic_relevance_rel}), and Google results are different among the ADD steps:
\begin{itemize}[leftmargin=10pt]
\item[--] For \textit{Identify design concepts} and \textit{Select design concepts}, topics on comparing solutions, and achieving quality attributes are most relevant. But Google cannot distinguish highly relevant topics, and retrieves low relevant topics.
%\item[--] For the \textit{Select design concepts}, topics on comparing solutions and achieving quality attributes are most relevant for practitioners; Google behaves best, but still retrieves topics of low relevance.
\item[--] For \textit{Instantiate architecture elements}, topics on elaborating and evaluating solutions such as patterns, and how to design are most relevant; Google retrieves articles on patterns, but misses other highly relevant topics.
\end{itemize}
%$\bullet$ Google results are different among the ADD steps:
%\begin{itemize}[leftmargin=10pt]
%\item[--] For the \textit{Identify design concepts}, Google cannot distinguish highly relevant topics.
%\item[--] For the \textit{Select design concepts}, Google behaves best, but still retrieves topics of low relevance.
%\item[--] For the \textit{Instantiate architecture elements}, Google is biased for articles on patterns, and misses other highly relevant topics.
%\end{itemize}
\end{tcolorbox}
\vspace{-4mm}
\section {Discussion}
\label{sec:discussion}
\vspace{-2mm}
\subsection {Implications for researchers}
\label{sec:discussionResearchers}
\textit{The results of RQ1} split architectural blogs into types as retrieved by Google. Researchers could use these types and dataset to develop specialized Web search approaches, which classify architectural blogs based on their types. For instance, an approach can allow practitioners to filter personal blogs from other types of architectural blogs, to facilitate searching for AK, because some types of blogs are specialized in certain architectural topics (see Figure \ref{fig:Topics-co}). Furthermore, the types of blogs and our dataset facilitate future studies on AK. For instance, researchers recently analyze grey literature (including blog articles) to extract design decisions (e.g. on Microservices APIs \cite{Singjai2021PractitionerTheory}). Using the types of blogs and our dataset, researchers can collect a sample of architectural blog articles from specific types of blogs to explore certain topics.

\textit{The results of RQ2} help to understand the similarities and differences between the AK in architectural blogs and AK in Stack Overflow. In Table \ref{tab:AKComparision}, we compare the AK in architectural blogs based on our results (see Section \ref{sec:RQ2Results}), and the AK in Stack Overflow based on Soliman et al. \cite{Soliman2016}. We could infer the following similarities and differences:

\noindent $\bullet$ The majority of articles in architectural blogs discuss patterns and components, while the majority of architectural posts in Stack Overflow discuss technologies and features.

\noindent $\bullet$ Most of the articles in architectural blogs and architectural posts in Stack Overflow evaluate and compare solutions.

\noindent $\bullet$ Architectural blogs involve articles that guide software engineers to design a system (i.e. How to design). These are complex articles with multiple steps and issues. On the other hand, posts in Stack Overflow focus on a single issue, and do not involve multiple steps or issues to design a system.

Based on the comparison between architectural blogs and Stack Overflow, researchers could benefit from both sources. For example, an approach can find AK on component design from blogs, and AK on technologies from Stack Overflow.

The results of co-occurrences between LDA topics and qualitative topics (see Section \ref{sec:RQ2ResultsCooccurence}) show that using LDA in combination with AK ontology (see Section \ref{sec:RQ2Steps}) identifies reasonable architectural topics that correspond and complement the qualitative topics. Thus, researchers could re-use this approach (LDA + AK ontology) to explore architectural topics in other sources (e.g. issue trackers \cite{Bhat2017AutomaticApproach} and mailing lists \cite{Fu2021ADecisions}).

\textit{The results of RQ3} can guide researchers to develop heuristics, which improve the effectiveness of AK search approaches. For instance, Soliman et al. \cite{Soliman2018ImprovingCommunities} developed a heuristic-based specialized search approach to find relevant AK in Stack Overflow. %, the approach shows its significance compared to traditional search approaches. 
Similarly, using the results of RQ3, researchers can develop heuristics to improve the search for AK in blogs by promoting highly relevant topics, and filtering low relevant topics. For example, when practitioners search for AK that pertains to the \textit{identify design concepts} an approach can promote topics \textit{Compare solutions} and \textit{Achieve quality attributes using component design}, and filter out topics \textit{Analyze decision factors} and \textit{Elaborate and evaluate a solution}. 

%because these articles have the highest relevance for this design step. Moreover, search approaches can filter-out articles from low relevant topics such as \textit{Analyze decision factors} and \textit{Elaborate and evaluate a solution}.

\begin{table}[]
\caption{Comparison of AK in architectural blogs (see Section \ref{sec:RQ2Results}) and Stack Overflow architectural posts \cite{Soliman2016}.}
\begin{tabular}{p{1.55cm} p{3.1cm} p{4.1cm}}
\hline
\multicolumn{1}{l}{} & \multicolumn{1}{c}{\textbf{Architectural blogs}} & \multicolumn{1}{c}{\textbf{Stack Overflow}}
\\ \hline
\textbf{Types of \newline architectural solutions} & 46\% components and patterns, 38\% technologies, 16\% decision factors & 78\% technologies and their features, 22\% architectural components (incl. combinations with other solutions)
\\ \hline
\textbf{Purpose of \newline article \newline or post} & 48\% elaborate, evaluate, compare solutions, 20\% list solutions, 32\% How to design or implement      & 50.7\% evaluate solutions (incl. comparison), 40.7\% synthesize solutions (incl. list solutions), 8.6\% combine evaluation and synthesis 
\\ \hline
\end{tabular}
\label{tab:AKComparision}
\vspace{-6mm}
\end{table}
\vspace{-1mm}
\subsection {Implications for practitioners}
\label{sec:discussionPractitioners}
\textit{The results of RQ1} inform practitioners about the types of architectural blogs, which could help them to share and find AK on the Web. On the one hand, practitioners could decide on suitable types of blogs to share their AK. For instance, practitioners should better share AK in community blogs, because Web search engines find them better than other types. On the other hand, the types of blogs and our dataset can guide practitioners to search for AK on the Web. For instance, practitioners should go directly to specific websites for certain types of architectural blogs in our dataset, and search in these websites directly, because some types of blogs are not well retrieved by Web search engines (e.g. personal blogs).

\textit{The results of RQ2} inform practitioners about the topics in architectural blogs. This can help practitioners to search for relevant articles in architectural blogs. For instance, practitioners could use terms and AK concepts of the LDA topics (see Table \ref{tab:LDATopics}) to search for specific architectural topics using keywords searches. Furthermore, the co-occurrences between blog types and topics (see Section \ref{sec:RQ2ResultsCooccurence}) guide practitioners to find certain architectural topics. For example, practitioners could search directly in community blogs (e.g. dzone.com) to find articles about architectural patterns and component design, because community blogs significantly co-occur with the LDA topic \textit{Design using patterns and components} (see Figure \ref{fig:Topics-co}). 
%On the other hand, practitioners should better browse articles in personal blogs for topics regarding architectural implementation, because personal blogs significantly co-occur with the LDA topic \textit{Implement technologies and component design}, as well as the qualitative topic on \textit{How to implement} (see Figure \ref{fig:Topics-co}).

\textit{The results of RQ3} guide practitioners on topics of architectural blogs, which are relevant to each of the ADD steps. Thus, if a practitioner requires AK in order to perform a specific ADD step, she can search specifically for the topics that are mostly relevant to this step. For example, regarding the \textit{instantiate architecture elements} ADD step, practitioners can search for blog articles with topics \textit{design using patterns and components} and \textit{achieve quality attributes using component design}, because they have the highest relevance to this step.
\vspace{-1mm}
\section{Threats to validity}
\label{sec:threats}
\subsubsection{Construct validity}
One threat to construct validity is regarding the use of AK ontology \cite{Soliman2017DevelopingCommunities}, which map AK concepts into specific terms. The terms associated with each AK concept might not be complete, and thus might miss AK concepts in blog articles. Nevertheless, we followed an iterative process (see Section \ref{sec:RQ2Steps}) to enrich the AK ontology with new terms from architectural blogs to mitigate this threat.
\subsubsection{Reliability}
To identify types and topics of architectural blogs, we manually analyzed articles using grounded theory. This might involve a threat to the reliability of the study. Nevertheless, we carefully followed the steps of grounded theory, and discussed articles to ensure agreement among researchers. Finally, we measured the agreement among researchers (see Section \ref{sec:RQ1Agreement}), which indicated good agreement beyond chance. 
Regarding the LDA analysis, we provide the scripts and AK ontology online \cite{OnlineMaterial} to facilitate replicating the study.
% Say something about the disagreements
\subsubsection{External validity}
One threat to the external validity is the limited number of analyzed architectural blog articles (i.e. 718 articles), which might not generalize to all architectural blog articles on the Web. Nevertheless, our sample is significant \cite{NeuendorfTheGuidebook} with 95\% confidence level and 3.58\% error margin. Moreover, other studies that explored AK analyzed samples with comparable sizes: 858 Stack Overflow posts \cite{Soliman2016}, 781 issues from issue trackers \cite{Bhat2017AutomaticApproach}, and 980 decisions from mailing list \cite{Li2020AutomaticList}. Thus, our results provide a first hypothesis of AK in blogs, which can be well compared to related work.
%Another threat to external validity comes from the dataset of Soliman et al., which involve data from six design tasks that correspond to the three ADD design steps.
\section{Related Work}
\label{sec:relatedwork}
We are not aware of any dedicated studies on architectural blogs and their contained AK. Thus, our study is the first study focusing on architectural blogs. In this section, we discuss related work in the AK field, architectural grey-literature reviews, and studies on blogs in software engineering.

\subsubsection{Architectural knowledge}
Previous studies identified and modeled AK concepts on design decisions \cite{Jansen2005}, their kinds \cite{Kruchten2006}, their reasoning \cite{Tang2007a}, and solutions alternatives \cite{Zimmermann2009}. Although these studies identified AK concepts, they do not explore concrete sources of AK.
Recent efforts explored AK in different software repositories and on the Web. For example, Gorton et al.  \cite{GortonICSA2017} found AK regarding architectural tactics in technology documentation, and Bhat et al. \cite{Bhat2017AutomaticApproach} found AK regarding types of decisions in issue tracking systems.
%Moreover, Soliman et al. \cite{Soliman2021AnSystems} explored different AK concepts, their variants and co-occurrences in issue tracking systems.
%Soliman et al. \cite{Soliman2018ImprovingCommunities} improved the effectiveness of searching for AK in Stackoverflow. The approach uses machine learning and heuristics to re-rank the results of search engines.
Furthermore, Bi et al. \cite{Bi2021MiningOverflow} explored quality attributes and architectural tactics in Stack Overflow.
Recently, Fu et al. \cite{Fu2021ADecisions} developed an approach to identify decisions in mailing list. Moreover, Mahadi et al. \cite{Mahadi2022ConclusionDiscussions} developed an approach to identify design discussions in pull requests.
However, all previously mentioned approaches do not explore architectural blogs and their contained AK.
\subsubsection{Architectural grey-literature reviews}
Grey-literature may involve blog articles but also other sources (e.g. forums). Researchers recently utilized grey literature to answer research questions. For example, Soldani et al. \cite{Soldani2018TheReview} analyzed grey-literature to explore the benefits and drawbacks of microservices. While, Singjai et al. \cite{Singjai2021PractitionerTheory} analyzed grey-literature to determine relationships between Microservice APIs and Domain-Driven Design. However, these studies extract certain information (e.g. decision rules \cite{Singjai2021PractitionerTheory}) from blogs, but do not explore architectural blogs and its AK as one source of AK.
\subsubsection{Blogs in software engineering}
Blogs did not have much attention from software engineering researchers. Pagano et al. \cite{WalidBlogs} was the first to analyze blog articles, and determined topics discussed by software developers such as features and domain concepts. Similarly, Parnin et al. \cite{Parnin2013BloggingDirections} analyzed topics of software blogs (e.g. documentation, technology discussion), and surveyed practitioners to determine motivations and challenges to write blog articles.
Recently, Williams et al. \cite{Williams2019HowPosts,Williams2021TowardsArticles} developed an approach to assess the credibility of blog articles.
However, these studies analyzed blogs with different focus, and did not explore architectural blogs and AK.
\vspace{-1mm}
\section{Conclusion and Future Work}
\label{sec:conclusion}
We aimed to explore architectural blogs, their types, topics and relevance to design steps. To this end, we analyzed architectural blog articles using qualitative and quantitative research methods. Our results show that architectural blog articles are shared in different types with different hosting and authorship policies, and discuss different topics that involve listing, elaborating, and evaluating solutions such as patterns, components and technologies. Furthermore, we found that some topics are more relevant for practitioners to make certain design steps than others. However, Web search engines (e.g. Google) do not always retrieve the most relevant topics for design steps. Our future work aims to develop approaches that automatically identify and classify blogs to improve the search for architectural knowledge in blogs.

%\bibliographystyle{IEEEtran}
%\bibliography{references}

\bibliographystyle{IEEEtran}
\bibliography{references}

\begin{thebibliography}{10}
\providecommand{\url}[1]{#1}
\csname url@samestyle\endcsname
\providecommand{\newblock}{\relax}
\providecommand{\bibinfo}[2]{#2}
\providecommand{\BIBentrySTDinterwordspacing}{\spaceskip=0pt\relax}
\providecommand{\BIBentryALTinterwordstretchfactor}{4}
\providecommand{\BIBentryALTinterwordspacing}{\spaceskip=\fontdimen2\font plus
\BIBentryALTinterwordstretchfactor\fontdimen3\font minus
  \fontdimen4\font\relax}
\providecommand{\BIBforeignlanguage}[2]{{%
\expandafter\ifx\csname l@#1\endcsname\relax
\typeout{** WARNING: IEEEtran.bst: No hyphenation pattern has been}%
\typeout{** loaded for the language `#1'. Using the pattern for}%
\typeout{** the default language instead.}%
\else
\language=\csname l@#1\endcsname
\fi
#2}}
\providecommand{\BIBdecl}{\relax}
\BIBdecl

\bibitem{Elaborateandevaluateasolution1}
\BIBentryALTinterwordspacing
Evaluating message brokers – amazon sqs. [Online]. Available:
  \url{https://neiljbrown.com/2017/08/26/evaluating-message-brokers-amazon-sqs/}
\BIBentrySTDinterwordspacing

\bibitem{Elaborateandevaluateasolution2}
\BIBentryALTinterwordspacing
Microservices architecture: What, when, and how. [Online]. Available:
  \url{https://dzone.com/articles/microservices-architecture-what-when-how}
\BIBentrySTDinterwordspacing

\bibitem{Elaborateandevaluateasolution4}
\BIBentryALTinterwordspacing
Components involved in creating a robust micro service architecture. [Online].
  Available:
  \url{https://www.javacodegeeks.com/2016/05/components-involved-creating-robust-micro-service-architecture.html}
\BIBentrySTDinterwordspacing

\bibitem{Elaborateandevaluateasolution3}
\BIBentryALTinterwordspacing
Why is loose coupling between services so important? [Online]. Available:
  \url{https://www.ben-morris.com/why-is-loose-coupling-between-services-so-important/}
\BIBentrySTDinterwordspacing

\bibitem{List2}
\BIBentryALTinterwordspacing
Best 6 stock market apis for 2022. [Online]. Available:
  \url{http://www.columbia.edu/~tmd2142/best-6-stock-market-apis-for-2020.html}
\BIBentrySTDinterwordspacing

\bibitem{List3}
\BIBentryALTinterwordspacing
Design patterns for microservices. [Online]. Available:
  \url{https://ngkester123.medium.com/design-patterns-for-microservices-9340b72407bb}
\BIBentrySTDinterwordspacing

\bibitem{List6}
\BIBentryALTinterwordspacing
11 ways to improve json performance \& usage. [Online]. Available:
  \url{https://stackify.com/top-11-json-performance-usage-tips/}
\BIBentrySTDinterwordspacing

\bibitem{List4}
\BIBentryALTinterwordspacing
5 fundamentals to a successful microservice design. [Online]. Available:
  \url{https://techbeacon.com/app-dev-testing/5-fundamentals-successful-microservice-design}
\BIBentrySTDinterwordspacing

\bibitem{List5}
\BIBentryALTinterwordspacing
Best json library for java. [Online]. Available:
  \url{https://www.javaguides.net/2019/07/best-json-library-for-java.html}
\BIBentrySTDinterwordspacing

\bibitem{List1}
\BIBentryALTinterwordspacing
What are 10 microservices best practices? [Online]. Available:
  \url{https://www.devteam.space/blog/10-best-practices-for-building-a-microservice-architecture/}
\BIBentrySTDinterwordspacing

\bibitem{Comparesolutions1}
\BIBentryALTinterwordspacing
Performance comparison of several common json libraries in java. [Online].
  Available: \url{https://www.programmersought.com/article/8810377980/}
\BIBentrySTDinterwordspacing

\bibitem{Comparesolutions2}
\BIBentryALTinterwordspacing
Synchronous vs. asynchronous request handling. [Online]. Available:
  \url{https://blog.trigent.com/synchronous-vs-asynchronous-request-handling/}
\BIBentrySTDinterwordspacing

\bibitem{Howtodesign3}
\BIBentryALTinterwordspacing
Rabbitchat - chat server in python using tornado, sockjs \& rabbitmq - also
  talking about tcp sockets, websockets \& python gil. [Online]. Available:
  \url{https://medium.com/@anirbanroydas/rabbitchat-chat-server-using-tornado-sockjs-rabbitmq-also-talking-about-tcp-sockets-914a79bbcd2f}
\BIBentrySTDinterwordspacing

\bibitem{Howtodesign2}
\BIBentryALTinterwordspacing
Rest api best practices — design examples from java and spring web services.
  [Online]. Available:
  \url{https://dzone.com/articles/rest-api-best-practices-with-design-examples-from}
\BIBentrySTDinterwordspacing

\bibitem{Howtodesign4}
\BIBentryALTinterwordspacing
Pattern: Decompose by business capability. [Online]. Available:
  \url{https://microservices.io/patterns/decomposition/decompose-by-business-capability.html}
\BIBentrySTDinterwordspacing

\bibitem{Howtodesign5}
\BIBentryALTinterwordspacing
A web api design methodology. [Online]. Available:
  \url{https://www.infoq.com/articles/web-api-design-methodology/}
\BIBentrySTDinterwordspacing

\bibitem{Howtodesign6}
\BIBentryALTinterwordspacing
Routing topologies for performance and scalability with rabbitmq. [Online].
  Available:
  \url{https://spring.io/blog/2011/04/01/routing-topologies-for-performance-and-scalability-with-rabbitmq/}
\BIBentrySTDinterwordspacing

\bibitem{Howtoimplement3}
\BIBentryALTinterwordspacing
Microservices with spring. [Online]. Available:
  \url{https://spring.io/blog/2015/07/14/microservices-with-spring}
\BIBentrySTDinterwordspacing

\bibitem{Howtoimplement2}
\BIBentryALTinterwordspacing
Consuming a restful web service. [Online]. Available:
  \url{https://openliberty.io/guides/rest-client-java.html}
\BIBentrySTDinterwordspacing

\bibitem{Howtoimplement1}
\BIBentryALTinterwordspacing
Integrate with dynamics crm data using apache camel2. [Online]. Available:
  \url{https://www.cdata.com/kb/tech/dynamicscrm-jdbc-apache-camel.rst}
\BIBentrySTDinterwordspacing

\bibitem{Designusingpatternsandcomponents1}
\BIBentryALTinterwordspacing
How to design a web application: Software architecture 101. [Online].
  Available:
  \url{https://www.educative.io/blog/how-to-design-a-web-application-software-architecture-101}
\BIBentrySTDinterwordspacing

\bibitem{Designusingpatternsandcomponents2}
\BIBentryALTinterwordspacing
Design patterns for microservices. [Online]. Available:
  \url{https://dzone.com/articles/design-patterns-for-microservices-1}
\BIBentrySTDinterwordspacing

\bibitem{Achievequalityattributesusingcomponentsdesign1}
\BIBentryALTinterwordspacing
Kafka performance tuning — ways for kafka optimization. [Online]. Available:
  \url{https://medium.com/@rinu.gour123/kafka-performance-tuning-ways-for-kafka-optimization-fdee5b19505b}
\BIBentrySTDinterwordspacing

\bibitem{Achievequalityattributesusingcomponentsdesign2}
\BIBentryALTinterwordspacing
Which service: Rabbitmq vs apache kafka. [Online]. Available:
  \url{https://www.cloudkarafka.com/blog/which-service-rabbitmq-vs-apache-kafka.html}
\BIBentrySTDinterwordspacing

\bibitem{Implementtechnologiesandcomponentsdesign1}
\BIBentryALTinterwordspacing
Microservices with spring. [Online]. Available:
  \url{https://spring.io/blog/2015/07/14/microservices-with-spring}
\BIBentrySTDinterwordspacing

\bibitem{Implementtechnologiesandcomponentsdesign2}
\BIBentryALTinterwordspacing
Event-driven integration on kubernetes with camel \& keda. [Online]. Available:
  \url{https://tomd.xyz/kubernetes-event-driven-keda/}
\BIBentrySTDinterwordspacing

\bibitem{Analyzedecisionfactors1}
\BIBentryALTinterwordspacing
Etl vs. elt and the benefits of data transformation in the cloud. [Online].
  Available:
  \url{https://solutionsreview.com/data-integration/etl-vs-elt-and-the-benefits-of-data-transformation-in-the-cloud/}
\BIBentrySTDinterwordspacing

\bibitem{Analyzedecisionfactors2}
\BIBentryALTinterwordspacing
Power apps and sap middleware integration to enhance the supply chain paradigm.
  [Online]. Available:
  \url{https://www.softwebsolutions.com/resources/powerapps-integration-with-SAP.html}
\BIBentrySTDinterwordspacing

\bibitem{Compareandevaluatetechnologies1}
\BIBentryALTinterwordspacing
Activemq vs rabbitmq: Key features and benefits. [Online]. Available:
  \url{https://www.openlogic.com/blog/activemq-vs-rabbitmq}
\BIBentrySTDinterwordspacing

\bibitem{Designusingmultipletechnologies1}
\BIBentryALTinterwordspacing
Advanced architecture for asp.net core web api. [Online]. Available:
  \url{https://www.infoq.com/articles/advanced-architecture-aspnet-core/}
\BIBentrySTDinterwordspacing

\end{thebibliography}


\begin{thebibliography}{10}
\providecommand{\url}[1]{#1}
\csname url@samestyle\endcsname
\providecommand{\newblock}{\relax}
\providecommand{\bibinfo}[2]{#2}
\providecommand{\BIBentrySTDinterwordspacing}{\spaceskip=0pt\relax}
\providecommand{\BIBentryALTinterwordstretchfactor}{4}
\providecommand{\BIBentryALTinterwordspacing}{\spaceskip=\fontdimen2\font plus
\BIBentryALTinterwordstretchfactor\fontdimen3\font minus
  \fontdimen4\font\relax}
\providecommand{\BIBforeignlanguage}[2]{{%
\expandafter\ifx\csname l@#1\endcsname\relax
\typeout{** WARNING: IEEEtran.bst: No hyphenation pattern has been}%
\typeout{** loaded for the language `#1'. Using the pattern for}%
\typeout{** the default language instead.}%
\else
\language=\csname l@#1\endcsname
\fi
#2}}
\providecommand{\BIBdecl}{\relax}
\BIBdecl

\bibitem{Kruchten2006}
P.~Kruchten, P.~Lago, and H.~van Vliet, ``{Building Up and Reasoning About
  Architectural Knowledge},'' in \emph{Quality of Software Architectures}, vol.
  4214.\hskip 1em plus 0.5em minus 0.4em\relax Springer Berlin Heidelberg,
  2006, pp. 43--58.

\bibitem{Jansen2005}
A.~Jansen and J.~Bosch, ``{Software Architecture as a Set of Architectural
  Design Decisions},'' in \emph{WICSA}, 2005, pp. 109--120.

\bibitem{Soliman2021ExploringKnowledge}
M.~Soliman, M.~Wiese, Y.~Li, M.~Riebisch, and P.~Avgeriou, ``{Exploring Web
  Search Engines to Find Architectural Knowledge},'' in \emph{2021 IEEE 18th
  International Conference on Software Architecture (ICSA)}.\hskip 1em plus
  0.5em minus 0.4em\relax IEEE, 3 2021, pp. 162--172.

\bibitem{BuschmannHenneySchmidt07a}
F.~Buschmann, K.~Henney, and D.~C. Schmidt, \emph{{Pattern-Oriented Software
  Architecture, Volume 4: A Pattern Language for Distributed Computing}}.\hskip
  1em plus 0.5em minus 0.4em\relax Chichester, UK: Wiley, 2007.

\bibitem{SolimanWicsa2015}
M.~Soliman, M.~Riebisch, and U.~Zdun, ``{Enriching Architecture Knowledge with
  Technology Design Decisions},'' in \emph{Working International Conf. on
  Software Architecture}, 5 2015.

\bibitem{Tang2007a}
A.~Tang, Y.~Jin, and J.~Han, ``{A rationale-based architecture model for design
  traceability and reasoning},'' \emph{Journal of Systems and Software},
  vol.~80, no.~6, pp. 918--934, 6 2007.

\bibitem{Soliman2018ImprovingCommunities}
M.~Soliman, A.~R. Salama, M.~Galster, O.~Zimmermann, and M.~Riebisch,
  ``{Improving the Search for Architecture Knowledge in Online Developer
  Communities},'' in \emph{International Conference on Software Architecture
  2018}.\hskip 1em plus 0.5em minus 0.4em\relax IEEE Inc., 7 2018.

\bibitem{KazmanDesigningSoftware2016}
R.~Kazman and H.~Cervantes, \emph{{Designing Software Architectures: A
  Practical Approach}}.\hskip 1em plus 0.5em minus 0.4em\relax Addison-Wesley
  Professional, 2016.

\bibitem{Manteuffel2018AnProjects}
C.~Manteuffel, P.~Avgeriou, and R.~Hamberg, ``{An exploratory case study on
  reusing architecture decisions in software-intensive system projects},''
  \emph{Journal of Systems and Software}, vol. 144, pp. 60--83, 10 2018.

\bibitem{Hassan2008TheRepositories}
A.~E. Hassan, ``{The road ahead for mining software repositories},''
  \emph{Proceedings of the 2008 Frontiers of Software Maintenance, FoSM 2008},
  pp. 48--57, 2008.

\bibitem{Bhat2017AutomaticApproach}
M.~Bhat, K.~Shumaiev, A.~Biesdorf, U.~Hohenstein, and F.~Matthes, ``{Automatic
  extraction of design decisions from issue management systems: A machine
  learning based approach},'' in \emph{Lecture Notes in Computer Science}, vol.
  10475 LNCS.\hskip 1em plus 0.5em minus 0.4em\relax Springer Verlag, 2017, pp.
  138--154.

\bibitem{Li2020AutomaticList}
X.~Li, P.~Liang, and Z.~Li, ``{Automatic Identification of Decisions from the
  Hibernate Developer Mailing List},'' \emph{ACM International Conference on
  the Evaluation and Assessment in Software Engineering (EASE '20)}, pp.
  51--60, 4 2020.

\bibitem{Bi2021MiningOverflow}
T.~Bi, P.~Liang, A.~Tang, and X.~Xia, ``{Mining architecture tactics and
  quality attributes knowledge in Stack Overflow},'' \emph{Journal of Systems
  and Software}, p. 111005, 5 2021.

\bibitem{GortonICSA2017}
I.~Gorton, R.~Xu, Y.~Yang, H.~Liu, and G.~Zheng, ``{Experiments in Curation:
  Towards Machine-Assisted Construction of Software Architecture Knowledge
  Bases},'' in \emph{IEEE/IFIP ICSA 2017}, 4 2017, pp. 79--88.

\bibitem{Soliman2017DevelopingCommunities}
\BIBentryALTinterwordspacing
M.~Soliman, M.~Galster, and M.~Riebisch, ``{Developing an Ontology for
  Architecture Knowledge from Developer Communities},'' \emph{International
  Conference on Software Architecture}, 5 2017. [Online]. Available:
  \url{https://doi.org/10.1109/ICSA.2017.31}
\BIBentrySTDinterwordspacing

\bibitem{WalidBlogs}
D.~Pagano and W.~Maalej, ``{How do open source communities blog?}''
  \emph{Empirical Software Engineering}, vol.~18, no.~6, pp. 1090--1124, 2013.

\bibitem{Williams2017TowardResearch}
A.~Williams and A.~Rainer, ``{Toward the use of blog articles as a source of
  evidence for software engineering research},'' \emph{Proceedings of the 21st
  International Conference on Evaluation and Assessment in Software Engineering
  (EASE'17)}, vol. Part F1286, pp. 280--285, 6 2017.

\bibitem{Silva2021TopicResearch}
\BIBentryALTinterwordspacing
C.~C. Silva, M.~Galster, and F.~Gilson, ``{Topic modeling in software
  engineering research},'' \emph{Empirical Software Engineering 2021 26:6},
  vol.~26, no.~6, 9 2021. [Online]. Available:
  \url{https://link.springer.com/article/10.1007/s10664-021-10026-0}
\BIBentrySTDinterwordspacing

\bibitem{OnlineMaterial}
\BIBentryALTinterwordspacing
``{Online material}.'' [Online]. Available:
  \url{https://github.com/m-a-m-s/Explore-Architectural-Blogs}
\BIBentrySTDinterwordspacing

\bibitem{Strauss1990BasicsTechniques.}
A.~Strauss and J.~M. Corbin, \emph{{Basics of qualitative research: Grounded
  theory procedures and techniques.}}\hskip 1em plus 0.5em minus 0.4em\relax
  Thousand Oaks, CA, US: Sage Publications, Inc, 1990.

\bibitem{Stol2016GroundedGuidelines}
K.~J. Stol, P.~Ralph, and B.~Fitzgerald, ``{Grounded theory in software
  engineering research: A critical review and guidelines},'' in
  \emph{Proceedings - International Conference on Software Engineering}, vol.
  14-22-May-2016.\hskip 1em plus 0.5em minus 0.4em\relax IEEE Computer Society,
  5 2016, pp. 120--131.

\bibitem{PearsonSquare}
K.~Pearson, ``{On a criterion that a given system of deviations from the
  probable in the case of correlated system of variables is such that it can be
  reasonably supposed to have arisen from random sampling},'' pp. 157--175,
  1900.

\bibitem{Kruskal1952UseAnalysis}
W.~H. Kruskal and W.~A. Wallis, ``{Use of Ranks in One-Criterion Variance
  Analysis},'' \emph{Journal of the American Statistical Association}, vol.~47,
  no. 260, pp. 583--621, 1952.

\bibitem{Fisher1992StatisticalWorkers}
\BIBentryALTinterwordspacing
R.~A. Fisher, ``{Statistical Methods for Research Workers},'' pp. 66--70, 1992.
  [Online]. Available:
  \url{https://link.springer.com/chapter/10.1007/978-1-4612-4380-9_6}
\BIBentrySTDinterwordspacing

\bibitem{Singjai2021PractitionerTheory}
A.~Singjai, U.~Zdun, and O.~Zimmermann, ``{Practitioner Views on the
  Interrelation of Microservice APIs and Domain-Driven Design: A Grey
  Literature Study Based on Grounded Theory},'' \emph{Proceedings - IEEE 18th
  International Conference on Software Architecture, ICSA 2021}, pp. 25--35, 3
  2021.

\bibitem{Soliman2016}
M.~Soliman, M.~Galster, A.~R. Salama, and M.~Riebisch, ``{Architectural
  knowledge for technology decisions in developer communities: An exploratory
  study with StackOverflow},'' \emph{Working International Conference on
  Software Architecture (WICSA)}, pp. 128--133, 7 2016.

\bibitem{Fu2021ADecisions}
L.~Fu, P.~Liang, X.~Li, and C.~Yang, ``{A machine learning based ensemble
  method for automatic multiclass classification of decisions},'' \emph{ACM
  International Conference on Evaluation and Assessment in Software Engineering
  (EASE 2021)}, pp. 40--49, 6 2021.

\bibitem{NeuendorfTheGuidebook}
K.~A. Neuendorf, \emph{{The Content Analysis Guidebook}}, 2nd~ed.\hskip 1em
  plus 0.5em minus 0.4em\relax SAGE Publications.

\bibitem{Zimmermann2009}
O.~Zimmermann, J.~Koehler, F.~Leymann, R.~Polley, and N.~Schuster, ``{Managing
  architectural decision models with dependency relations, integrity
  constraints, and production rules},'' \emph{Journal of Systems and Software},
  vol.~82, no.~8, pp. 1249--1267, 2009.

\bibitem{Mahadi2022ConclusionDiscussions}
\BIBentryALTinterwordspacing
A.~Mahadi, N.~A. Ernst, and K.~Tongay, ``{Conclusion stability for natural
  language based mining of design discussions},'' \emph{Empirical Software
  Engineering}, 1 2022. [Online]. Available:
  \url{https://link.springer.com/article/10.1007/s10664-021-10009-1}
\BIBentrySTDinterwordspacing

\bibitem{Soldani2018TheReview}
J.~Soldani, D.~A. Tamburri, and W.~J. Van Den~Heuvel, ``{The pains and gains of
  microservices: A Systematic grey literature review},'' \emph{Journal of
  Systems and Software}, vol. 146, pp. 215--232, 12 2018.

\bibitem{Parnin2013BloggingDirections}
C.~Parnin, C.~Treude, and M.~A. Storey, ``{Blogging developer knowledge:
  Motivations, challenges, and future directions},'' \emph{IEEE International
  Conference on Program Comprehension}, pp. 211--214, 2013.

\bibitem{Williams2019HowPosts}
A.~Williams and A.~Rainer, ``{How do empirical software engineering researchers
  assess the credibility of practitioner–generated blog posts?}'' \emph{ACM
  International Conference on the Evaluation and Assessment on Software
  Engineering (EASE 2019)}, pp. 211--220, 4 2019.

\bibitem{Williams2021TowardsArticles}
A.~Williams, M.~Shardlow, and A.~Rainer, ``{Towards a corpus for credibility
  assessment in software practitioner blog articles},'' \emph{ACM International
  Conference on Evaluation and Assessment in Software Engineering (EASE 2021)},
  pp. 100--108, 6 2021.

\end{thebibliography}

\bibliographystyleB{IEEEtran}
\bibliographyB{blogs} 

\end{document}